\documentclass[twocolumn,aps,prc,eqsecnum,preprintnumbers,showpacs,
showkeys,nofootinbib,superscriptaddress,fleqn,floatfix,tightenlines, 10pt]
{revtex4-1}
\usepackage{amsmath}
\usepackage{epsfig}
\usepackage{color}
\usepackage{txfonts} 
\usepackage{dcolumn} 
\everymath{\displaystyle}

\begin{document}
\title{Tractable \boldmath{$T$}-matix model for reaction processes in
muon catalyzed fusion
\boldmath{ $(dt\mu)_{J=v=0} \to   \; \alpha + n + \mu + 17.6\, {\rm MeV}
\; \mbox{or} \; (\alpha \mu)_{nl} + n +17.6 \,{\rm MeV}$  }
}

\author{Qian Wu}
~\email{qwu@nju.edu.cn}
\affiliation{School of Physics, Nanjing University, Nanjing 21000, China}
\affiliation{Institute of Modern Physics, Chinese Academy of Sciences, Lanzhou 730000, China}

\author{Masayasu Kamimura}
\email{mkamimura@a.riken.jp}
\affiliation{Meson Science Laboratory, RIKEN Nishina Center, RIKEN, Wako 351-0198, Japan}
\parindent=12pt

\begin{abstract}
Reaction processes in muon catalyzed fusion ($\mu$CF),
$(dt\mu)_{J=v=0} \to \alpha + n + \mu + 17.6\,{\rm MeV}\:$
or $ \;(\alpha \mu)_{nl} + n + 17.6\,{\rm MeV}$ in the D-T mixture
was comprehensively studied by Kamimura, Kino and Yamashita
[Phys. Rev. C {\bf 107}, 034607 (2023)] by solving 
the $dt\mu$-$\alpha n\mu$
coupled channel (CC) Schr\"odinger equation under a boundary condition where
the muonic molecule $(dt\mu)_{J=v=0}$ was set as the initial state 
and the outgoing wave was \mbox{in the $\alpha n\mu$ channel.}
\mbox{We approximate} this  CC framework
and propose a considerably more tractable model
using the $T$-matrix method based on the Lippmann-Schwinger 
equation.  Nuclear interactions adopted in the $T$-matrix 
model are determined by reproducing the  cross section  of the  reaction  
$d + t \to \alpha + n + 17.6\,{\rm MeV}$ at low energies.
The cross section of the strong-coupling rearrangement reaction  
is presented in a simple closed form based on our new model.
This \mbox{$T$-matrix} model have reproduced most of the calculated 
results on the above $\mu$CF reaction reported by Kamimura \mbox{{\it et al.}} 
(2023) and is applicable to other $\mu$CF systems such as
$(dd\mu)$, $(tt\mu)$, $(dt\mu)^*$,  $(dd\mu)^*$.
\end{abstract}
\maketitle

\section{INTRODUCTION}
\begin{figure} [b!]
\begin{center}
\epsfig{file=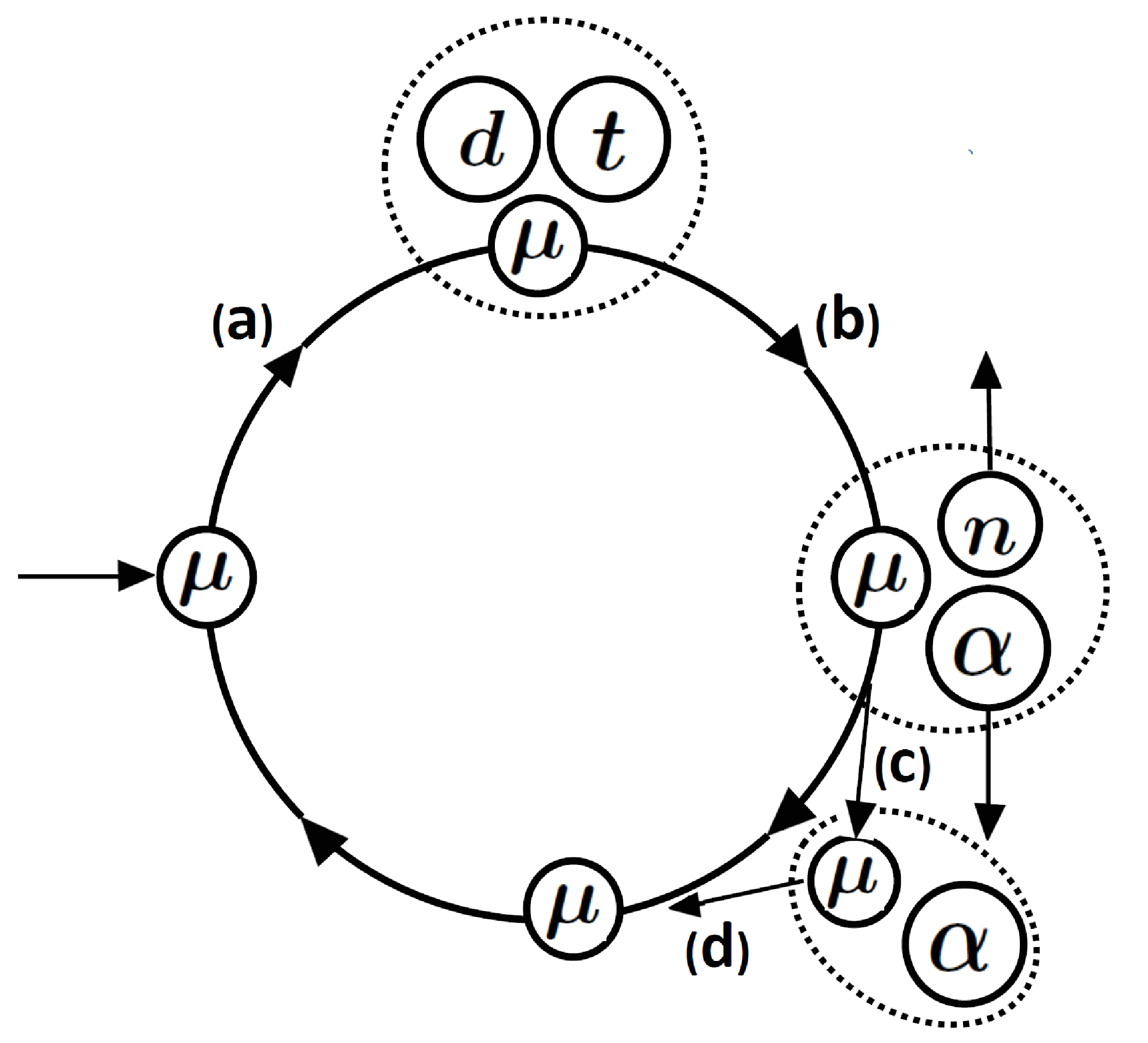,width=3.8cm}
\caption{
Schematic diagram of the $\mu$CF cycle by a muon injected into
the D/T mixture.
(a) Formation of $dt\mu$ molecule, (b) fusion reaction,
(c) $\alpha$-$\mu$ initial sticking, and (d) muon reactivation.
This illustration is taken from Ref.~\cite{Kamimura:2021msf}.
}
\label{fig:cycle}
\end{center}
\end{figure}

To ensure that nuclear fusion occurs within nuclear distance (a few fms)
the Coulomb barrier between two nuclei must be overcome, which typically
requires very high temperatures. 
At low temperatures, negative muons injected into a mixture of
deuterium (D) and tritium (T) can catalyze the fusion reaction
\begin{equation}
 \qquad \;\;    d+t\rightarrow\alpha+n+17.6\; {\rm MeV}  ,
    \label{eq:1dtan}
\end{equation}
which is energetically the most effective nuclear fusion reaction. 
Following the catalyzed  reaction, free muons can
facilitate another fusion reaction 
taking the well-known  cycle illustrated in Fig.~\ref{fig:cycle}.
This cyclic reaction is called muon catalyzed fusion ($\mu$CF). 
The fusion of the muonic molecule $dt\mu$ has attracted more attention
compared to that of molecules such as $dd\mu$, $tt\mu$ 
from the perspective  of utilization as a future energy source. 
The history of $\mu$CF since 1947 has been reviewed 
in Refs.~~\cite{Breunlich:1989vg,Ponomarev:1990pn,Kamimura1998anp,Froelich1992}.
The present status of the study of $dt\mu$ fusion is briefly summarized 
in Ref.~\cite{Kamimura:2021msf}.

Recently, the $\mu$CF has regained  considerable attention
owing to experimental and theoretical developments 
i) in the production of energy by $\mu$CF using the high-temperature
{\it gas} target of a D/T mixture with high thermal efficiency and
ii) in  an ultra-slow negative muon
beam by utilizing  $\mu$CF
for various applications including  scanning negative muon microscope
and an injection source for the muon collider. 
This is explained in detail
in the Introduction of Ref.~\cite{Kamimura:2021msf}.

The study of Ref.~\cite{Kamimura:2021msf} performed
a comprehensive examination of the following $dt\mu$ fusion reaction:
\begin{subequations}
\label{eq:mucf-reaction}
\begin{align}
(dt\mu)_{J=v=0} & \to \;\alpha \, + n \, + \mu  + 17.6 \,\mbox{MeV}
\label{first equation}  \\
      & \searrow    \; (\alpha \mu)_{nl} + n  + 17.6 \,\mbox{MeV}.
\label{second equation} 
\end{align}
\end{subequations}
\noindent  
For the first time, this study solved a coupled-channel (CC) Schr\"odinger equation
for the reaction (\ref{eq:mucf-reaction}) using the appropriate
boundary conditions where $(dt\mu)_{J=v=0}$ was set as the initial state
and the outgoing wave was expressed in the $\alpha n\mu$ channel.
All interactions were  selected
such that the low-energy cross sections of the reaction (\ref{eq:1dtan})
were reproduced using the CC calculations for the reaction. 
They calculated the fusion rate of the $(dt\mu)_{J=v=0}$ molecule,
energy (momentum) spectra of the muon emitted by  $\mu$CF, and 
$\alpha$-$\mu$ sticking probability~\cite{Kamimura:2021msf}.

As investigated in Ref.~\cite{Yamashita:2022rtu}, the 
$dd\mu$ and $tt\mu$  fusions play important roles
for the  new kinematics of the $\mu$CF cycle
in case of high temperature D-T mixtures. 
Although the $dd\mu$ fusion is known to be
considerably weaker than  $dt\mu$ fusion, the former
should be studied more precisely 
because the $dd\mu$ experiment is an important preliminary experiment
of  $dt\mu$, which has a difficulty in the tritium treatment.
Although a study on reaction processes in the $dd\mu$
and $tt\mu$ fusion is underway~\cite{Kamimura2023x}  using the same CC method 
of Ref.~\cite{Kamimura:2021msf}, solving the 
CC equations for the reactions expressed as Eq.~(\ref{eq:mucf-reaction})
is difficult. 
Therefore, creating a starting point from the CC method~\cite{Kamimura:2021msf},  
such as an approximation method
that simulates their results more easily would be beneficial. 


The purpose of the present paper is to propose
considerably more tractable $T$-matrix model than the aforementioned CC model of 
Ref.~\cite{Kamimura:2021msf}. Instead of directly solving 
the CC Schr\"odinger equations
for the reactions (\ref{eq:1dtan}) and  (\ref{eq:mucf-reaction}),
we approximate the $T$-matrix based 
on the Lippmann-Schwinger equation~\cite{Lippmann} that is equivalent to
the Schr\"odinger equation.
Nuclear interactions are selected to reproduce  
the cross section of reaction (\ref{eq:1dtan})
at the center-of-mass (c.m.) energy $\sim \! 1\!-\!300$ keV. 
The total angular momentum $I$ and parity of the scattering
states is known to be $I^\pi=3/2^+$.
In the following, we explain the scenario of our model in Steps 
\mbox{i) -- v)}.

\vskip 0.1cm
Step i) First, we  reproduce the  cross section of low-energy
reaction (\ref{eq:1dtan}) by using the $d$-$t$ optical-potential 
\mbox{model adopted} in Refs.~\cite{KAMIMURA1989-BENCHMARK}. 
Absorption cross section
is regarded as the cross section of reaction (\ref{eq:1dtan})
because no other open channels \mbox{exist} at these energies 
than the $d$-$t$ and \mbox{$\alpha$-$n$} channels.
The $d$-$t$ scattering wave function is denoted as 
$\Phi_{dt, \frac{3}{2}M}^{\rm (opt)}$ 
(Sec.~II A). 

Step ii) We diagonalize the $dt\mu$ Hamiltonian composed of
the Coulomb potentials and  $d$-$t$ optical potential
to obtain the complex eigenenergy ($E_{\rm R}+i E_{\rm I}$) and
wave function $\Phi_{J=v=0}(dt\mu)$ of the ground state of
$(dt\mu)$ molecule. The fusion rate (decay rate) of the
molecule is expressed as $-2E_{\rm I}/\hbar$ (Sec.~II B).
Here, we introduce 
$\Phi^{(J=v=0)}_{\frac{3}{2} M}(dt\mu)$ as the product of 
$\Phi_{J=v=0}(dt\mu)$ and the $d$-$t$ spin function 
$\chi_{\frac{3}{2} M} (dt)$.

Step iii) The cross section of reaction (\ref{eq:1dtan})
can also be expressed by the exact $T$-matrix 
introduced in Eq.~(4.1) of Ref.~\cite{Kamimura:2021msf}, 
where $\Psi^{(+)}_\alpha$ denotes the exact
solution of the CC wave function for the reaction (\ref{eq:1dtan})
and $V_\beta$ stands for a coupling potential
$V_{dt, \alpha n}$  between the $d$-$t$ and $\alpha$-$n$ channels.
In our model,
we replace the exact $\Psi^{(+)}_\alpha$ with the $d$-$t$ wave function
$\Phi_{dt,\, \frac{3}{2}M}^{\rm (opt)}$ obtained in \mbox{Step i)}.
$\Phi_{dt,\, \frac{3}{2}M}^{\rm (opt)}$ is considered to
include the effects of the outgoing $\alpha$-$n$ channel
using the imaginary part of the $d$-$t$ potential. 
The coupling potential 
$V_{dt, \alpha n}$  is determined 
by reproducing the cross section of reaction (\ref{eq:1dtan}). 
Notably, the cross section of 
a strong coupling rearrangement reaction  as shown in  (1.1) 
is expressed, based on our model, in a simple closed form 
that can successfully reproduce the observed data at low energies (Sec.~III).
This coupling potential $V_{dt, \alpha n}$ is used in Steps iv) and v).

Step iv) In the work of Ref.~[5], the $T$-matrix (4.1) was used to study
the three-body fusion reaction (1.2) in the manner of (5.2)--(5.7)
with outgoing waves in the $(\alpha \mu)\!-\!n$ channel.
There, the exact $\Psi^{(+)}_\alpha$ in the $T$-matrix 
was replaced with 
the three-body CC wave function $\Psi^{(+)}_{\frac{3}{2} M}$ of Eq.(3.3) [5]. 
\mbox{In our model,} 
the \mbox{exact} $\Psi^{(+)}_\alpha$ is replaced 
with the wave function $\Phi^{(J=v=0)}_{\frac{3}{2} M}(dt\mu)$
of the $dt\mu$ molecule obtained in Step ii). 
Further, the \mbox{$\alpha$-$\mu$ sticking} probability
is derived using the fusion rates to the $\alpha$-$\mu$ continuum states 
and those to the $\alpha$-$\mu$ bound states. 

Step v) We make another calculation of the
fusion rate of  reaction (\ref{eq:mucf-reaction})
using  $T$-matrix (4.1) in Ref.~\cite{Kamimura:2021msf} 
with outgoing waves in the $(\alpha n)\!-\!\mu$ channel.
The exact $\Psi^{(+)}_\alpha$ of Eq.~(3.3) in Ref.~[5]
is again \mbox{replaced} 
with $\Phi^{(J=v=0)}_{\frac{3}{2} M}(dt\mu)$ as in iv). 
We further \mbox{calculate} the momentum and energy spectrum of 
the muons emitted by $\mu$CF.

We shall examine 20 sets of the parameters for the nuclear interactions 
and show that the results do not significantly depend on the choice of 
the parameter sets
as long as the reaction (\ref{eq:1dtan}) is explained by using them.
We shall report that most of the results obtained in 
Ref.~\cite{Kamimura:2021msf} are
well reproduced by the present model.

This paper is organized as follows: 
In Sec.~II, using the optical-potential model,
we calculate the cross section of reaction (1.1) and
fusion rate of the $(dt\mu)_{J=v=0}$ molecule.
In Sec.~III, the coupling potential between  $d$-$t$ and $\alpha$-$n$
channels is determined using the $T$-matrix method.
In Sec.~IV, we calculate the $\alpha$-$\mu$ sticking probability
and the fusion rate of $(dt\mu)_{J=v=0}$  with the
method described in Step iv). 
In Sec.~V,  the spectra of the muons emitted by $\mu$CF and
the fusion rate of the $(dt\mu)_{J=v=0}$ are calculated  using the
method in Step v).
The conclusions are presented  in Sec.~VI.

\section{Optical-potential model for fusion processes}

\subsection{Fusion cross section}

Following Step i), we first investigate the fusion rection (\ref{eq:1dtan})
by using the optical-potential model of Ref.~[9]. 
The potential parameters 
of the nuclear $d$-$t$ potential are determined by reproducing the cross section 
of the reaction.
The total angular-momentum and parity $J^\pi$ of the 
reaction (\ref{eq:1dtan}) at low energies is $I^\pi=3/2^+$ with $S$-wave 
and spin 3/2 in the $d$-$t$ channel while $D$-wave and spin 1/2 
in the $\alpha$-$n$ channel.

We present the $d$-$t$ scattering
wave function $\Phi_{dt,\,\frac{3}{2}M}^{\rm (opt)}(E, {\bf r})$ 
at the c.m. energy $E$
as 
\begin{eqnarray}
\Phi_{dt,\,\frac{3}{2}M}^{\rm (opt)}(E, {\bf r}) 
= \phi_{dt, 00}^{\rm (opt)}(E, {\bf r})  \, 
\chi_{\frac{3}{2}M}(dt) ,
\end{eqnarray}
where  $\phi_{dt, 00}^{\rm (opt)}(E, {\bf r})$ is the spatial part of the $S$-wave
function and $\chi_{\frac{3}{2} M}(dt)$ is the $d$-$t$ spin 3/2 function.
Schr\"{o}dinger equation for $\phi_{dt, 00}^{\rm (opt)}(E, {\bf r})$  is presented as
\begin{eqnarray}
\label{eq:opt-Sch-2body}
 &&   ( H_{dt} - E)\, \phi_{dt, 00}^{\rm (opt)} (E,{\bf r}) =0, \\
\label{eq:opt-Hamil}
 &&  H_{dt} = T_{{\bf r}}
             + V_{dt}^{{\rm (N})}(r) + iW_{dt}^{{\rm (N})}(r)
          + V_{dt}^{({\rm Coul})}(r),
\end{eqnarray}
where the spin-independent $d$-$t$ optical potential is given by
\begin{eqnarray}
\label{eq:opt-real}
   && \!  V_{dt}^{{\rm (N})}(r)=V_0/\{1+e^{(r-R_0)/a}\} , \\
\label{eq:opt-imag}
   && \!\!  W_{dt}^{{\rm (N})}(r)=W_0/\{1+e^{(r-R_{\rm I})/a_{\rm I}} \} , \\
\label{eq:opt-coul}
&& \!\!\!\!\! V_{dt}^{({\rm Coul})}(r)=\begin{cases}
           (e^2/(2R_{\rm c}))(3-r^2/R_{\rm c}^2)\,,
       &  \text{$ \; r < R_{\rm c}$},  \\
          e^2/r\,,  &  \text{$ \; r \geq  R_{\rm c}$}
\end{cases}
\end{eqnarray}
with taking the charge radius $R_{\rm c}=R_0$.  

\begin{figure}[h]  
\setlength{\abovecaptionskip}{0.cm}
\setlength{\belowcaptionskip}{-0.cm}
\centering
\includegraphics[width=0.46\textwidth]{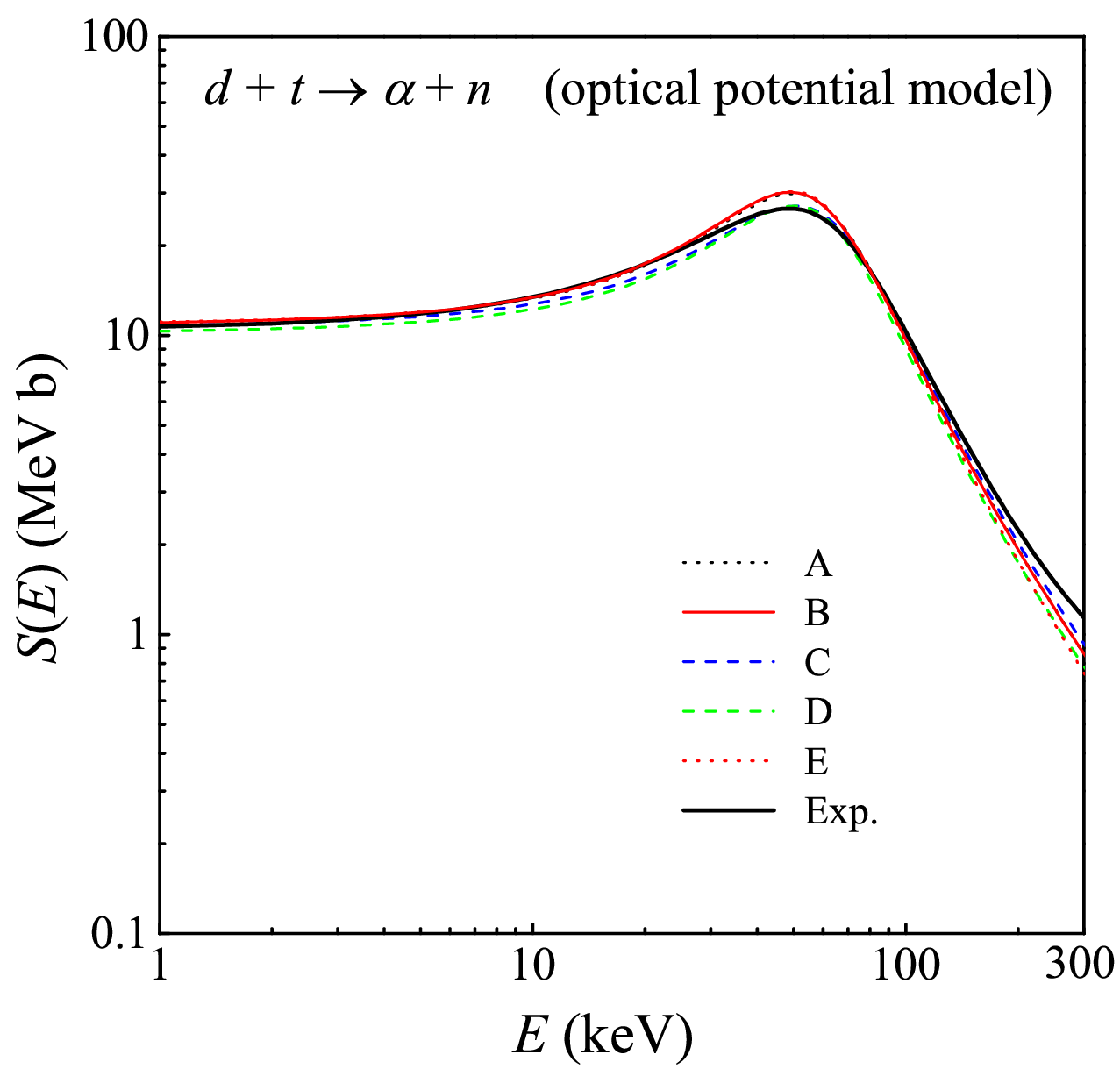}
\caption{Calculated $S$-factor $S(E)$ of the reaction 
$d+t \to \alpha + n + 17.6\,$ {\rm MeV}
using five different $d$-$t$ optical potentials 
A to E listed in \mbox{Table \ref{tab:vdt}}. The black solid line
(Exp.) is taken from a review paper \cite{Serpico:2004gx}; \mbox{it fits} the literature
data using the function $S(E) = (26-0.361E + 248E^2 )/{1 + [(E-0.0479)/0.0392]^2}$ 
MeV b ($E$ in MeV). 
}
\label{fig:opm-S}
\end{figure}

In the energy regions shown in  Fig.~\ref{fig:opm-S},
only the $\alpha$-$n$ channel is open,
except for the incoming channel. 
Therefore, the absorption
cross section becomes the cross section of
the \mbox{reaction (\ref{eq:1dtan}) as}
\begin{eqnarray}
\sigma_{dt \to \alpha n}(E)=\frac{2I+1}{(2I_d+1)(2I_t+1)}
\frac{\pi}{k^2} (1-|S_J(E)|^2), \;\;
\end{eqnarray}
with  $S$-matrix $S_J(E)$. In Eq.~(2.7), $I_d=1$, $I_t=1/2$, and 
$k=\sqrt{2 \mu_{dt} E}/\hbar$.
The astrophysical $S$-factor $S(E)$ is derived from the cross section as
\begin{eqnarray}
\sigma_{dt \to \alpha n}(E)=S(E)\, e^{-2\pi \eta(E)}/E,
\label{eq:S-factor}
\end{eqnarray}
where $\eta(E)$ denotes the Sommerfeld parameter.

Owing to the lack of $d$-$t$ elastic scattering information for $E\lesssim$ 300 keV 
demonstrating the nuclear-interaction effect, it is impossible to determine 
a unique $d$-$t$ optical potential based on the observed 
$S$-factor in Fig.~\ref{fig:opm-S}  (black solid line). 
Therefore, in Ref.~\cite{KAMIMURA1989-BENCHMARK}, 
five  sets of  $d$-$t$ optical potentials, denoted as A-E, 
were selected  to reproduce the observed data. 
In the present study, the parameter $W_0$ is changed slightly  to improve 
the agreement with the experimental $S$-factor $S(E)$ for $E\leq$ 10 keV. 
The calculated $S(E)$ values  with  optical potential A-E are 
shown in Fig.~\ref{fig:opm-S}; within the experimental error range [10],
the observed $S(E)$ is reproduced well.
The potential parameters are listed in Table.~\ref{tab:vdt}.

\begin{table}[b]
\centering
\caption{Five sets (A to E) of
   the $d$-$t$ optical-potential parameters.}
\begin{tabular}
{p{1.2cm}<{\centering}p{1.0cm}<{\centering}p{1.0cm}<{\centering}p{1.0cm}<{\centering}p{1.0cm}<{\centering}p{1.0cm}<{\centering}p{1.0cm}<{\centering}}
       \hline
       \hline
       \noalign{\vskip 0.1 true cm}
          & $V_0$ & $R_0$ & $a$& $W_0$ & $R_I$ & $a_I$ \\
          & (MeV) & (fm) & (fm) & (MeV) & (fm) & (fm) \\
          \noalign{\vskip 0.1 true cm}
        \hline
        \noalign{\vskip 0.1 true cm}
        A & 58.52 &2.5 &0.3 &$-0.30$ &2.5 &0.3 \\
        B & 38.01 &3.0 &0.5 &$-0.30$ &3.0 &0.5 \\
        C & 28.27 &3.0 &1.0 &$-0.66$ &2.0 &1.0 \\
        D & 16.04 &5.0 &0.3 &$-0.22$ &2.5 &0.3 \\
        E & 13.19 &5.0 &1.0 &$-0.33$ &3.0 &1.0 \\
        \noalign{\vskip 0.1 true cm}
        \hline
        \hline
    \end{tabular}
    \label{tab:vdt}
\end{table}
%

%
\subsection{Fusion rate of muonic molecule} 
\label{sec:gemdtm}

\begin{figure}[b]
\setlength{\abovecaptionskip}{0.cm}
\setlength{\belowcaptionskip}{-0.cm}
\centering
\includegraphics[width=0.46\textwidth]{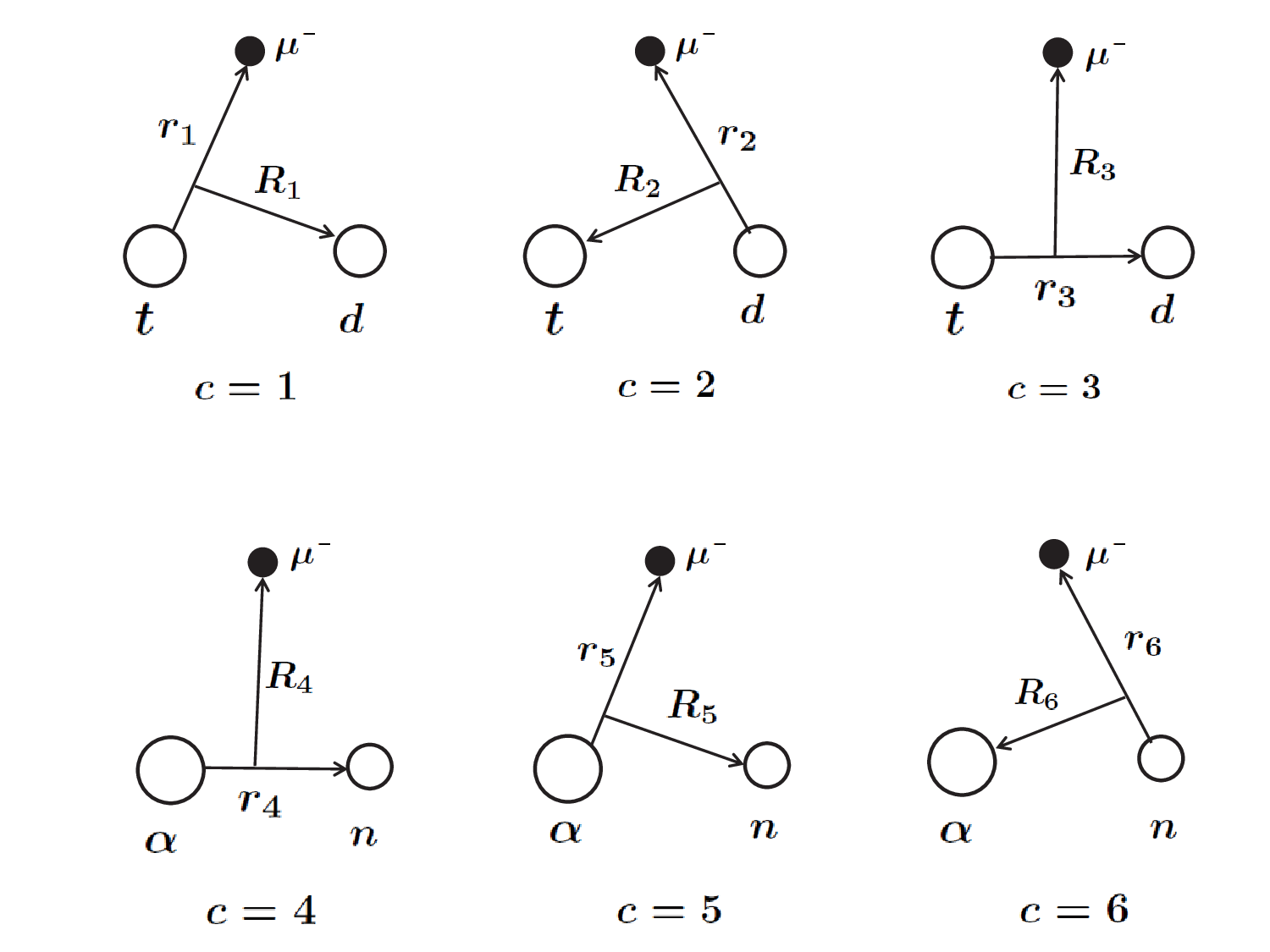}
\caption{Three-body Jacobi coordinates used in this study.}
\label{fig:jaco6}
\end{figure}
%

Following Step ii),
we calculate the fusion rate of the reaction (1.2)  by 
diagonalizing the $dt\mu$ three body Hamiltonian including the Coulomb force 
and the $d$-$t$ nuclear complex potentials  determined 
in the previous subsection. 
\mbox{We perform} a non-adiabatic three-body calculation of the ground-state
wave function  of the $dt\mu$ molecule, $\Phi_{J=v=0}(dt\mu)$,
using the Gaussian Expansion Method for 
few-body systems (GEM)~\cite{Kamimura:1988zz,Kameyama89,Hiyama03}:
\begin{eqnarray}
\label{eq:Sch-dtmu-bound}
   && ( H_{dt\mu} - E_{00})\, \Phi_{J=v=0}(dt\mu) =0,  \\
\label{eq:Hamil-dtmu-bound}
   &&  H_{dt\mu}=  T_{{\bf r}_c} +  T_{{\bf R}_c}
          + V^{({\rm C})}(r_1) + V^{({\rm C})}(r_2)        \nonumber \\
    && \qquad         + V^{({\rm N})}_{dt}(r_3) + iW^{({\rm N})}_{dt}(r_3)
         + V^{({\rm C})}_{dt}(r_3)\,.
\end{eqnarray}
$\Phi_{J=v=0}(dt\mu)$ is constructed 
as the sum of  
amplitudes of the three rearrangement channels c=1, 2, and 3 
as shown in Fig.~\ref{fig:jaco6}:
\begin{equation}
\!\!
{\Phi}_{J=v=0}(d t \mu)=\Phi_{0}^{(1)}\!\left(\mathbf{r}_1, \mathbf{R}_1\right)+
\Phi_{0}^{(2)}\!\left(\mathbf{r}_2, \mathbf{R}_2\right)+
\Phi_{0}^{(3)}\!\left(\mathbf{r}_3, \mathbf{R}_3\right).
\label{eq:gemwf}
\end{equation}
The amplitude of each channel $c$ is expanded in terms of Gaussian basis 
functions of the Jacobian coordinates ${\rm r}_c$ \mbox{and ${\rm R}_c$:}
\begin{equation}
\Phi_{0}^{(c)}\left(\mathbf{r}_c, \mathbf{R}_c\right)=
\!\!\sum_{n_s l_c, N_c L_c} A_{n_c l_c, N_c L_s}^{(c)}
\left[\phi_{n_c l_c}\left(\mathbf{r}_c\right) \psi_{N_c L_c}
\left(\mathbf{R}_c\right)\right]_{00},
\end{equation}
where $c=1-3$ and
\begin{equation}
\begin{aligned}
& \phi_{n l m}(\mathbf{r})=\phi_{n l}(r) Y_{l m}(\hat{\mathbf{r}}), \\
& \phi_{n l}(r)=N_{n l} r^l e^{-v_n r^2}, \quad\left(n=1-n_{\max }\right), \\
& \psi_{N L M}(\mathbf{R})=\psi_{N L}(R) Y_{L M}(\widehat{\mathbf{R}}), \\
& \psi_{N L}(R)=N_{N L} R^L e^{-\lambda_N R^2}, \quad\left(N=1-N_{\max }\right),
\end{aligned}
\label{eq:gaussian-basis}
\end{equation}
with  normalization constants $N_{n l}$ and $N_{N L}$.
%
The Gaussian range parameters $\nu_n$ and $\lambda_n$ are chosen 
in geometric progression.
\begin{equation}
\begin{aligned}
& v_n=1 / r_n^2, \quad r_n=r_1 a^{n-1},\left(n=1-n_{\max }\right), \\
& \lambda_N=1 / R_N^2, \quad R_N=R_1 A^{N-1},\left(N=1-N_{\max }\right) .
\label{eq:gaussian-range}
\end{aligned}
\end{equation}
Subsequently, the eigenenergy and  wave function are obtained 
using the Rayleigh-Ritz variational method.
The advantages of using the GEM basis functions are explained in detail
in Sec.~III~A in Ref.~\cite{arxiv2112.08399v1}.

\begin{table}[t]
\centering
\caption{Nonlinear variational parameters of the Gaussian basis functions 
in Eqs.~(\ref{eq:gaussian-basis})--(\ref{eq:gaussian-range}). 
$r_1$($R_1$) and $r_{\rm max}$($R_{\rm max}$) 
are in units of $a_\mu$ where $a_{\mu}=\hbar^2/m_\mu e^2= 255.9$ fm, 
$m_\mu$ being the muon mass.}
\begin{tabular}{p{0.5cm}<{\centering}p{0.5cm}<{\centering}p{0.5cm}
<{\centering}p{1.0cm}<{\centering}p{1.0cm}<{\centering}p{0.5cm}
<{\centering}p{1.0cm}<{\centering}p{1.0cm}<{\centering}p{1.0cm}<{\centering}}
\noalign{\vskip 0.1 true cm}
\hline \hline 
\noalign{\vskip 0.1 true cm}$\mathrm{c}$ & $l_c$ & $n_{\max }$ & $\begin{array}{c}r_1 \\
{[a_{\mu}]}\end{array}$ & $\begin{array}{c}r_{n_{\max }} \\
{[a_{\mu}]}\end{array}$ & $L_c$ & $N_{\max }$ & $\begin{array}{c}R_1 \\
{[a_{\mu}]}\end{array}$ & $\begin{array}{c}R_{N_{\max }} \\
{[a_{\mu}]}\end{array}$ \\
\noalign{\vskip 0.1 true cm}
\hline
\noalign{\vskip 0.1 true cm}
1 & 0 & 25 & 0.1 & 10 & 0 & 15 & 0.05 & 10 \\
2 & 0 & 25 & 0.1 & 10 & 0 & 15 & 0.05 & 10 \\
3 & 0 & 25 & 0.1 & 10 & 0 & 15 & 0.05 & 10 \\
1 & 1 & 15 & 0.2 & 10 & 1 & 15 & 0.3 & 15 \\
2 & 1 & 15 & 0.2 & 10 & 1 & 15 & 0.3 & 15 \\
3 & 1 & 15 & 0.2 & 10 & 1 & 15 & 0.3 & 15 \\
3 & 0 & 25 & 0.001 & 0.05 & 0 & 15 & 0.05 & 10 \\
\noalign{\vskip 0.1 true cm}
\hline
\hline
\end{tabular}
\label{tab:gaussian-para}
\end{table}
\begin{table}[h]
\centering
   \caption{Fusion rate of the reaction (\ref{eq:mucf-reaction})  
using the five sets (A to E) of
the $d$-$t$ optical-potential parameters in Table \ref{tab:vdt}.}
\begin{tabular}
{p{2.2cm}<{\centering}p{1.0cm}<{\centering}p{1.0cm}<{\centering}p{1.0cm}<{\centering}p{1.0cm}<{\centering}p{1.0cm}<{\centering}}
       \hline
       \hline
       \noalign{\vskip 0.1 true cm}
          & A & B & C & D & E \\
        \hline
        \noalign{\vskip 0.1 true cm}
        $\lambda^{(0)}_f(10^{12}{\rm s}^{-1})$ & 1.14 & 1.15 & 1.12 & 1.07 & 1.07\\
        \noalign{\vskip 0.1 true cm}
        \hline
        \hline
    \end{tabular}
    \label{tab:lambda-opm}
\end{table}

As the eigenenergy $E_{00}$ is a complex number, we write
$E_{00}=E_{00}^{(\rm real)}+i E_{00}^{(\rm imag)}$.  
We introduce $\varepsilon_{00}=E_{00}^{(\rm real)}-E_{\rm th}$, with 
$E_{\rm th}(=-2711.24$ eV) being the $(t\mu)_{1s}+d$ threshold energy. 
The diagonalization in the   
cases of $l_{\rm max}=4$ and $l_{\rm max}=1$ 
yield, respectively,  $\varepsilon_{00}=-319.14$ eV and $-319.12$ eV.
According to Ref.~\cite{Kamimura:2021msf}, 
the digits below 1 eV did not affect the reaction calculation. 
Thus, we employ $l_{\rm max}=1$. The input Gaussian basis is shown in 
Table~\ref{tab:gaussian-para}.
We took 7 lines of Gaussian basis parameters 
where the final line is effective 
to the $d$-$t$ nuclear interaction.

The fusion rate $\lambda^{(0)}_{\rm f}$ of reaction  (\ref{eq:mucf-reaction}) 
can be derived by 
$\lambda^{(0)}_{\rm f}=-2E_{00}^{({\rm imag})}/\hbar$ and is
given as $\lambda^{(0)}_{\rm f}=(1.11 \pm 0.04) \times 10^{12} {\rm s}^{-1}$
using five sets of the $d$-$t$ potentials
as presented in Table~\ref{tab:lambda-opm}.
It is consistent with the results 
$1.15 \times 10^{12} {\rm s}^{-1}$ obtained in Ref.~\cite{Kamimura:2021msf}.

We multiply $\Phi_{J=v=0}(dt\mu)$ by the $d$-$t$ 
spin function $\chi_{\frac{3}{2} M} (dt)$:
\begin{eqnarray}
\Phi^{(J=v=0)}_{\frac{3}{2} M}(dt\mu) = \Phi_{J=v=0}(dt\mu)\, 
\chi_{\frac{3}{2} M} (dt), 
\end{eqnarray}
which will be used in
Secs.~IV and V as the ground-state wave function the $(dt\mu)$ molecule.

\section{ {\boldmath $T$}-matix model for 
\mbox{\boldmath {$\lowercase{d + t} \to \alpha + \lowercase{n}
+ 17.6 $}}      \:\mbox{M\lowercase{e}V}    
}

In this section, following Step iii), we propose 
tractable $T$-matrix model for the reaction (1.1)
to approximate the CC model described in Ref.~[5]
(cf. Eqs.~\mbox{(2.1)-(2.10)})
using the results of the optical-potential model
in the previous section.

At low energies, the $d$-$t$ wave function has the total angular momentum 
$I=3/2$ with $S$-wave in the \mbox{$d$-$t$} channel and $D$-wave 
in the $\alpha$-$n$ channel.
The authors of Ref.~\cite{Kamimura:2021msf} solved the following $dt$-$\alpha n$
CC Schr\"{o}dinger equation ($Q=17.6 $ MeV) using the coordinates 
${\bf r}_3$ and ${\bf r}_4$ for the $d$-$t$ and $\alpha$-$n$ motion,
respectively (cf. Fig.3): 
\begin{eqnarray}
  \!\!\!\! \!\!\!\!    ( H_{dt} - E )\, \Phi_{dt,\, \frac{3}{2}M}({\bf r}_3)
  = -V^{({\rm T})}_{dt, \alpha n}({\bf r}_3, {\bf r}_4) \:  
\Phi^{(+)}_{\alpha n, \, \frac{3}{2}M}({\bf r}_4),\!\! && \nonumber \\
 \!\!\!\!\!\! \!\!\!\! \!\!\!\!   \left( H_{\alpha n} - (E + Q) \right)  
\Phi^{(+)}_{\alpha n, \, \frac{3}{2}M}({\bf r}_4)
   = -V^{({\rm T})}_{\alpha n, dt}({\bf r}_4, {\bf r}_3) 
\:\Phi_{dt, \, \frac{3}{2}M}({\bf r}_3)  &&
\label{sch-dt}
\end{eqnarray}
with trivial notations.
The channel-coupling potential $V^{({\rm T})}_{\alpha n, dt}$
($V^{({\rm T})}_{dt, \alpha n}$)
is of the tensor type.
The reaction cross section is presented as
\begin{eqnarray}
\sigma_{dt \to \alpha n}(E)=\frac{2I+1}{(2I_d+1)(2I_t+1)}
\frac{\pi}{k^2}\, |S_2^{(dt,\alpha n)}|^2 ,
\label{eq:sigma-CC-2body}
\end{eqnarray}
where the $S$-matrix $S_2^{(dt,\alpha n)}$ appears in the asymptotic 
form of the outgoing wave, $\Phi^{(+)}_{\alpha n,\, \frac{3}{2}M}$ [5].

According to the Lippmann-Schwinger theory,
the cross section is alternatively expressed exactly 
as following using the solution $\Phi_{dt, \,\frac{3}{2}M}$ and 
$ \Phi^{(+)}_{\alpha n, \,\frac{3}{2}M}$ of Eq.~(3.1)
(cf. Sec.~IV A of Ref.~\cite{Kamimura:2021msf}):  
\begin{eqnarray}
\label{eq:sigma-Tmat-CC-2body-exact}
&\!\!\!\! \sigma_{dt \to \alpha n}(E)= \frac{v_4}{v_3}
      \left( \frac{{\mu_{r_4}}}{2 \pi \hbar^2} \right)^2
\sum_{m_s} \int |\, T_{m_s}^{(1)} +  T_{m_s}^{(2)}\, |^2\, {\rm d}
{\bf \widehat{K}}, \quad \\
\label{eq:Tmat-(1)-CC-2body}
&\!\!\!\!\! T_{m_s}^{(1)}=
\langle\,e^{  i {\bf K} \cdot {\bf r}_4 }\,
       \chi_{\frac{1}{2} m_s} (n)  \,
    |\,  V^{({\rm T})}_{\alpha n, dt} \,| \,  \Phi_{dt, \,\frac{3}{2}M} \, \rangle,  \\
\label{eq:Tmat-(2)-CC-2body}
&\!\!\!\!\!\!  T_{m_s}^{(2)}=
\langle\,e^{  i {\bf K} \cdot {\bf r}_4 }\,
       \chi_{\frac{1}{2} m_s} (n)  \,
    |\,  V_{\alpha n} \,| \,   \Phi^{(+)}_{\alpha n, \,\frac{3}{2}M} \, \rangle,
\end{eqnarray}
where $\chi_{\frac{1}{2} m_s}(n)$ is the neutron spin function,
$V_{\alpha n}$ is the \mbox{$\alpha$-$n$} potential, 
$v_3$ ($v_4$) is the velocity of the $d$-$t$ ($\alpha$-$n$) relative motion
along ${\bf r}_3$ (${\bf r}_4$), $\mu_{r_4}$ is the reduced mass associated with
${\bf r}_4$, and ${\bf K}$ is  wave number vector. 

The coupling potential $V^{({\rm T})}_{\alpha n, dt}$ 
between the $\alpha n$-$d t$ channels
are taken in the following form~\cite{Kamimura:2021msf}:
\begin{eqnarray}
\label{eq:tensor-1}
\!\!\!\!\!\!\!\!\!\!\!
  V_{\alpha n, dt}^{({\rm T})}({\bf r}_4, {\bf r}_3)
 &\!\!\ =\!\!\!& v_0^{({\rm T})}r_{34}^2 e^{-\mu \,r_{34}^2- \mu' R_{34}^2}
        \,  \big[ Y_2({\widehat {\bf r}}_{34})\,
          {\cal S}_2(dt,\alpha n) \big]_{00},   
\end{eqnarray}
where  ${\bf r}_{34}= {\bf r}_3 - {\bf r_4}$ and
${\bf R}_{34}= {\bf r}_3 + {\bf r_4}$.
In Eq.~(\ref{eq:tensor-1}), ${\cal S}_2(dt,\alpha n)$ is a spin-tensor operator
comprising the spins of  $dt$- and \mbox{$\alpha n$-pairs.}
However,  the explicit form of ${\cal S}_2(dt,\alpha n)$ 
need not to be determined as explained below (cf. Eq.~(3.4)).

In our model, we perform another $T$-matrix calculation for 
$\sigma_{dt \to \alpha n}(E)$
by using Eq.~(\ref{eq:sigma-Tmat-CC-2body-exact}) 
considering the following approximation.
We replace $\Phi_{dt, \,\frac{3}{2}M}$ in  $T_{m_s}^{(1)}$
with $\Phi_{dt, \,\frac{3}{2}M}^{\rm (opt)}$
in Eq.~(2.1) and neglect $T_{m_s}^{(2)}$ because
$\Phi_{dt, \,\frac{3}{2}M}^{\rm (opt)}$ does not include $\alpha$-$n$ component explicitly.
However, $\Phi_{dt, \,\frac{3}{2}M}^{\rm (opt)}$ is  considered
to reflect the effect of the $\alpha$-$n$ channel 
using the imaginary potential $iW_{dt}^{({\rm N})}(r_3)$ in Eq.~(2.2).
Thus, Eqs.~(\ref{eq:sigma-Tmat-CC-2body-exact})--(\ref{eq:Tmat-(2)-CC-2body})
are approximated as
\begin{eqnarray}
\label{eq:sigma-Tmat-CC-2body}
 \!\!\!\!\! \!\!\!\!\! \!\! && \sigma^{\rm (our)}_{dt \to \alpha n}(E)
   = \frac{v_4}{v_3}
      \left( \frac{\mu_{r_4}}{2 \pi \hbar^2} \right)^2
\sum_{m_s} \int |\, T_{m_s}^{(1')} |^2\, {\rm d}
{\bf \widehat{K}}, \quad \\
\label{eq:Tmat-(1p)-CC-2body}
\!\!\!\!\! \!\!\!\!\! \!\! && T_{m_s}^{(1')}=
\langle\,e^{  i {\bf K} \cdot {\bf r}_4 }\,
       \chi_{\frac{1}{2} m_s} (n)  \,
    |\,  V^{({\rm T})}_{\alpha n, dt}({\bf r}_4, {\bf r}_3)
 \,| \,  \Phi_{dt, \,\frac{3}{2}M}^{\rm (opt)}(E,{\bf r}_3)
\, \rangle.
\end{eqnarray}
Instead of Eq.(\ref{eq:tensor-1}), we employ the following  
separable nonlocal form:
\begin{eqnarray}
\!\!\!\!\!\!\!\!\!
  V_{\alpha n, dt}^{({\rm T})}({\bf r}_4, {\bf r}_3)
 = v_0^{({\rm T})} r_4^2\,
          e^{-\mu_4 \,r_4^2- \mu_3 r_3^2} \,
  \big[ Y_2( {\widehat {\bf r}}_{4}  )\,
          {\cal S}_2(dt,\alpha n) \big]_{00},
\label{eq:tensor-2}
\end{eqnarray}
which is easier to handle  than Eq.~(\ref{eq:tensor-1}).
Consequently, the cross section~(\ref{eq:sigma-Tmat-CC-2body}) can be 
explicitly expressed  as
\begin{eqnarray}
 \sigma^{\rm (our)}_{dt \to \alpha n}(E) = \frac{v_4}{v_3}
      \left( \frac{\mu_{r_4}}{2 \pi \hbar^2} \right)^2
      \left| \,v_0^{(T)}S_0^{(T)} F_0\, J_2 \,\right|^2
\label{eq:sigma-answer}
\end{eqnarray}
with 
\begin{eqnarray}
&& \!\!\!\!\!\!\!\!\!\!\!\!\!\!\!\!\!\!\!\!\!\! F_0 = 
 \!\int \phi_{dt, 00}^{\rm (opt)}(E, {\bf r}_3) \,
e^{-\mu_3 r_3^2} \, {\rm d} {\bf r}_3, \\
&& \!\!\!\!\!\!\!\!\!\!\!\!\!\!\!\!\!\! \!\! \!\!
J_2 = 4\pi \!\!\int \!\! j_2(Kr_4) \, r_4^2 \,
e^{-\mu_4 r_4^2}\, r_4^2 \,{\rm d} r_4 
= \frac{1}{4} \left( \frac{\pi}{\mu_4} \right)^{\frac{3}{2}}\! 
\left( \frac{K}{\mu_4} \right)^{2} \! e^{-\frac{\mu_4 K^2}{4}}\!,
\end{eqnarray}
where $\phi_{dt, 00}^{\rm (opt)}(E, {\bf r}_3)$  is normalized 
asymptotically as   
\begin{eqnarray}
\!\!\!\!\!\!\!\!\! \phi_{dt, 00}^{\rm (opt)}(E, {\bf r}_3) 
 \stackrel{r_3 \to \infty}{\longrightarrow} e^{i \sigma_0}
\frac{F_0(k,r_3)}{kr_3}
 \,+\,  \mbox{(outgoing wave)} \quad
 \;\; 
\end{eqnarray}
with the S-wave Coulomb function $F_0(k,r)$ and phase shift $\sigma_0$.
$j_2(kr_4)$ is the spherical 
Bessel function of order 2. In Eq.~(\ref{eq:sigma-answer}), 
the constant $S_0^{({\rm T)}}$ is presented, independently of $m_s$, as
(cf. Eq.~(2.12) of Ref.~\cite{Kamimura:2021msf})
\begin{eqnarray}
\!\!\!\!\!\!\!  S^{{\rm (T)}}_0= \frac{1}{\sqrt{10}}
      \langle \,\chi_{\frac{1}{2} m_s}(\alpha n) \,|\,
 \big[ {\cal S}_2(dt,\alpha n) \,\chi_{\frac{3}{2}}(dt) \big]_{\frac{1}{2} m_s}
      \rangle \:.
\label{eq:spin-fac}
\end{eqnarray}
Therefore,  the explicit form of $S_2(dt,\alpha n)$ need not to be known, 
and $v_0^{{\rm (T)}}S_0^{{\rm (T)}}$ can be considered 
as an adjustable parameter for the $T$-matrix calculations.
%
\begin{table}[b]
\caption{Parameters of the $dt$-$\alpha n$ coupling potential
$V_{\alpha n, dt}^{({\rm T})}({\bf r}_4, {\bf r}_3)$ in Eq.~(\ref{eq:tensor-2}).
Sets A1-A4 were determined using the optical potential Set A in Table I;
similarly for the  others.
}
\label{tab:vdtan}
  \begin{tabular}{cccc}
\noalign{\vskip 0.1 true cm}
    \hline
    \hline
\noalign{\vskip 0.1 true cm}
potential set  &  $v_0^{({\rm T})} S_0^{({\rm T})}$  & $\;$  $\mu^{-1/2}_3$
& $\;$ $\;$ $\;$  $\mu^{-1/2}_4$ \\
\noalign{\vskip 0.1 true cm}
& $\;\;$  (MeV fm$^{-5}$)  & $\;\;$ (fm) $\;\;$ & $\;\;$$\;$ $\;$  (fm)$\;\;$ \\
\noalign{\vskip 0.1 true cm}
\noalign{\vskip 0.0 true cm}
    \hline
\noalign{\vskip 0.1 true cm}
   Set A1   & 2.307   & 1.6   & 1.6      \\
\noalign{\vskip 0.05 true cm}
   Set A2   & 0.138   & 2.0   & 5.2      \\
\noalign{\vskip 0.05 true cm}
   Set A3   & 0.245   & 3.2   & 1.6      \\
\noalign{\vskip 0.05true cm}
   Set A4   & 0.402   & 2.0   & 2.8      \\
\noalign{\vskip 0.05true cm}
\hline
\noalign{\vskip 0.05 true cm}
   Set B1   & 1.001   &  2.8   & 1.6     \\
\noalign{\vskip 0.05 true cm}
   Set B2   & 0.016   & 4.0   & 2.4      \\
\noalign{\vskip 0.05 true cm}
   Set B3   & 0.065   & 3.6   & 2.0      \\
\noalign{\vskip 0.05 true cm}
   Set B4   & 0.003   & 4.0   & 5.6      \\
\noalign{\vskip 0.05 true cm}
\hline
\noalign{\vskip 0.05true cm}
   Set C1   & 0.122   &  2.4   & 2.4     \\
\noalign{\vskip 0.05true cm}
   Set C2   & 0.052   & 1.6   & 6.4      \\
\noalign{\vskip 0.05true cm}
   Set C3   & 0.029   & 3.2   & 4.8      \\
\noalign{\vskip 0.05true cm}
   Set C4   & 0.191   & 4.4   & 1.6      \\
\noalign{\vskip 0.05true cm}
\hline
\noalign{\vskip 0.05true cm}
\noalign{\vskip 0.05true cm}
   Set D1   & 1.291   & 3.6   & 1.6      \\
\noalign{\vskip 0.05true cm}
   Set D2   & 0.151   & 3.2   & 2.0      \\
\noalign{\vskip 0.05true cm}
   Set D3   & 0.405   & 4.4   & 1.6      \\
\noalign{\vskip 0.05true cm}
   Set D4   & 0.032   & 4.0   & 5.6      \\
\noalign{\vskip 0.05true cm}
\hline
\noalign{\vskip 0.05true cm}
   Set E1   & 0.054   & 5.2   & 2.0      \\
\noalign{\vskip 0.05true cm}
   Set E2   & 0.166   & 2.0   & 2.0      \\
\noalign{\vskip 0.05true cm}
   Set E3   & 0.054   & 4.4   & 2.8      \\
\noalign{\vskip 0.05true cm}
   Set E4  & 0.019  & 4.4   & 5.2      \\
\noalign{\vskip 0.05true cm}
\hline
\hline
\noalign{\vskip 0.1 true cm}
\end{tabular}
\end{table}
%

\begin{figure}[h]
\setlength{\abovecaptionskip}{0.cm}
\setlength{\belowcaptionskip}{-0.cm}
\centering
\includegraphics[width=0.46\textwidth]{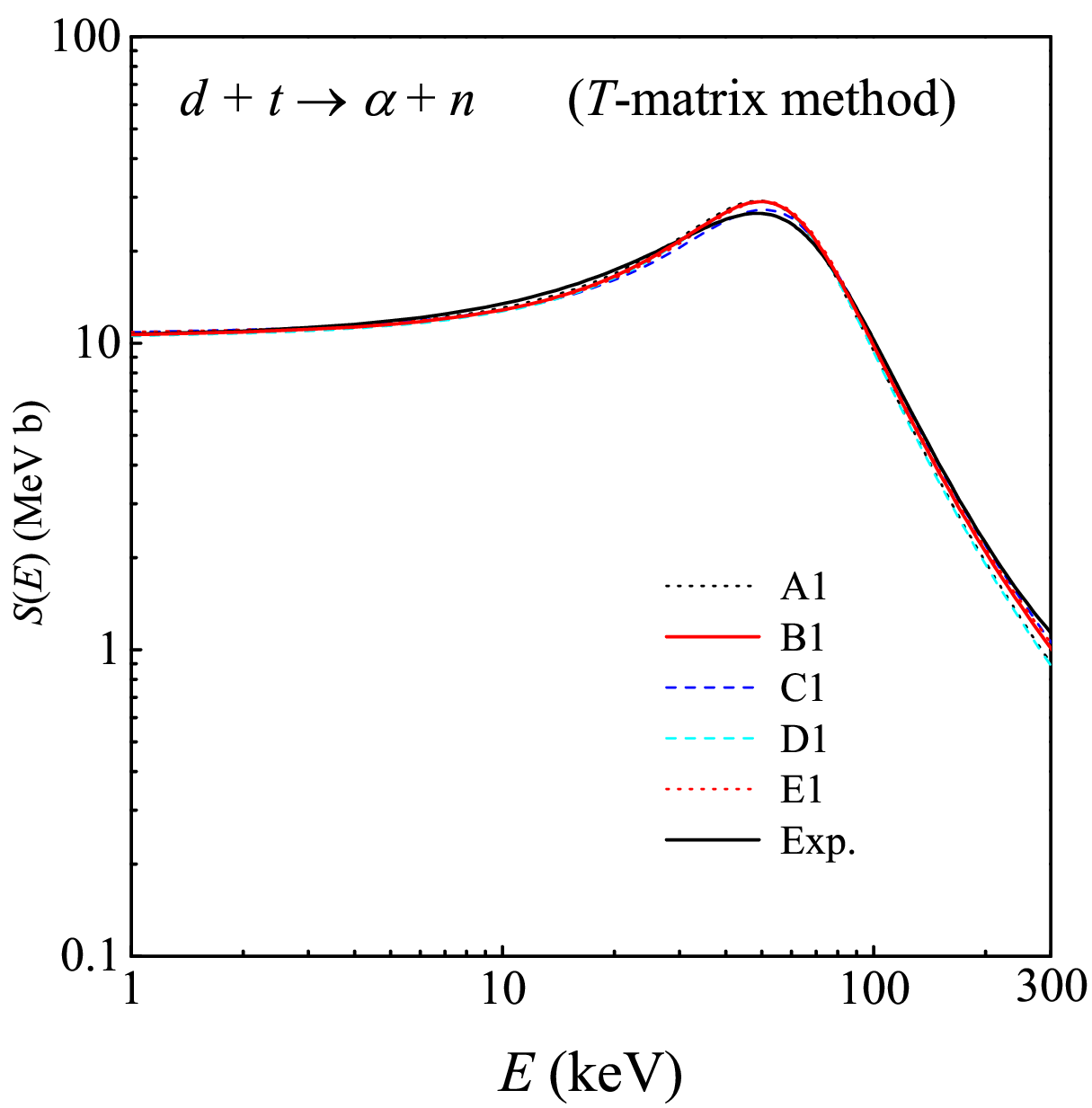}
\caption{Calculated $S$-factor $S(E)$ of the reaction 
$d+t \to \alpha + n + 17.6\,$ {\rm MeV}. Five lines A1-E1
denote the cases using the $dt$-$\alpha n$ coupling potentials
with the parameter Sets A1-E1, respectively, in \mbox{Table~IV.}
The black solid line indicating the experimental data is 
from a review paper \cite{Serpico:2004gx}.}
\label{fig:newT-S}
\end{figure}

We determined the parameter sets for the 
$dt$-$\alpha n$ coupling potential 
$V_{\alpha n, dt}^{({\rm T})}({\bf r}_4, {\bf r}_3)$ in Eq.~(3.9)
so as to reproduce the \mbox{observed} $S$-factor $S(E)$.
We employed, in Eq.~(3.8), 
$\Phi_{dt, \,\frac{3}{2}M}^{\rm (opt)}(E,{\bf r}_3)$ 
obtained in Sec.~II A  using the optical potentials A-E
in Table~I. The resulting parameters are listed in Table~IV.
Sets A1-A4 were obtained  using the optical potential of Set A, 
and similarly for others.

In Fig. \ref{fig:newT-S}, the calculated $S(E)$ factors 
using  Sets A1-E1 are illustrated. 
The experimental data are well reproduced 
with the same quality of fitting as in Fig.~4 by the CC calculation in of Ref.~[5]. 
The use of the other sets yielded a similar agreement.
Subsequently, we employ all \mbox{$dt$-$\alpha n$} tensor coupling potentials 
in Secs.~IV and V.  
The cross section of 
the strong coupling rearrangement reaction (1.1) 
is expressed  in a simple closed form (3.10)--(3.12)
that can reproduce observed data by tuning the 
parameters of $V_{\alpha n, dt}^{({\rm T})}({\bf r}_4, {\bf r}_3)$ ---- This
is a  key finding of this study.

Here, we highlight the consistency of our model in terms of its potentials.
In the  process of tuning the potential parameter sets, 
we considered the plane wave ($D$-wave) of the
$\alpha$-$n$ relative motion   
in the $T$-matrix (\ref{eq:Tmat-(1p)-CC-2body}) where 
the \mbox{$\alpha$-$n$} potential $V_{\alpha n}(r_4)$ did not appear
explicitly. We consider that
the effect of   $\alpha$-$n$ potential is effectively renormalized into
the coupling potential $V^{({\rm T})}_{\alpha n, dt}$
which is tuned to reproduce the observed  $S(E)$ without 
$V_{\alpha n}(r_4)$.
In Secs. IV and V, we  consistently use the same potentials 
without $V_{\alpha n}(r_4)$ in the $T$-matrix calculations 
of the three-body $dt\mu$-$\alpha n \mu$ system.
As shown later, most of the results of Ref.~\cite{Kamimura:2021msf} are 
well reproduced by our model.

%
\section{ {\boldmath $\alpha$-$\mu$} sticking probability}
\label{sec:stick}

After fusion occurs in the $dt\mu$ molecule, 
the emitted muon can be captured by the $\alpha$ particle or 
freely emitted, as shown in Eq.~(\ref{eq:mucf-reaction}).
In this section, according to Step iv) of the introduction,
we calculate the fusion rate of the reaction (\ref{eq:mucf-reaction})
and the $\alpha$-$\mu$ sticking probability  by
referring to the $T$-matrix calculation of those quantities 
in Sec.~IV of Ref.~\cite{Kamimura:2021msf} with the outgoing waves 
in channel $c=5$ (Fig.~\ref{fig:jaco6}).

The authors of Ref.~\cite{Kamimura:2021msf} solved the 
CC Schr{\" o}dinger equation (3.7) for the reaction (\ref{eq:mucf-reaction})
and generated the wave function $\Psi_{\frac{3}{2} M}^{(+)}(E)$ 
expressed as Eq.~(3.3).
They constructed the \mbox{$T$-matrix} elements (5.2)--(5.3)
for  use in another fusion calculation by substituting their  
$\Psi_{\frac{3}{2} M}^{(+)}(E)$ into the exact
wave function $\Psi_{\alpha}^{(+)}(E_\alpha)$  in the definition of 
$T$-matrix (4.1) in Ref.~\cite{Kamimura:2021msf}. 

\begin{figure}[b!]
\begin{center}
\epsfig{file=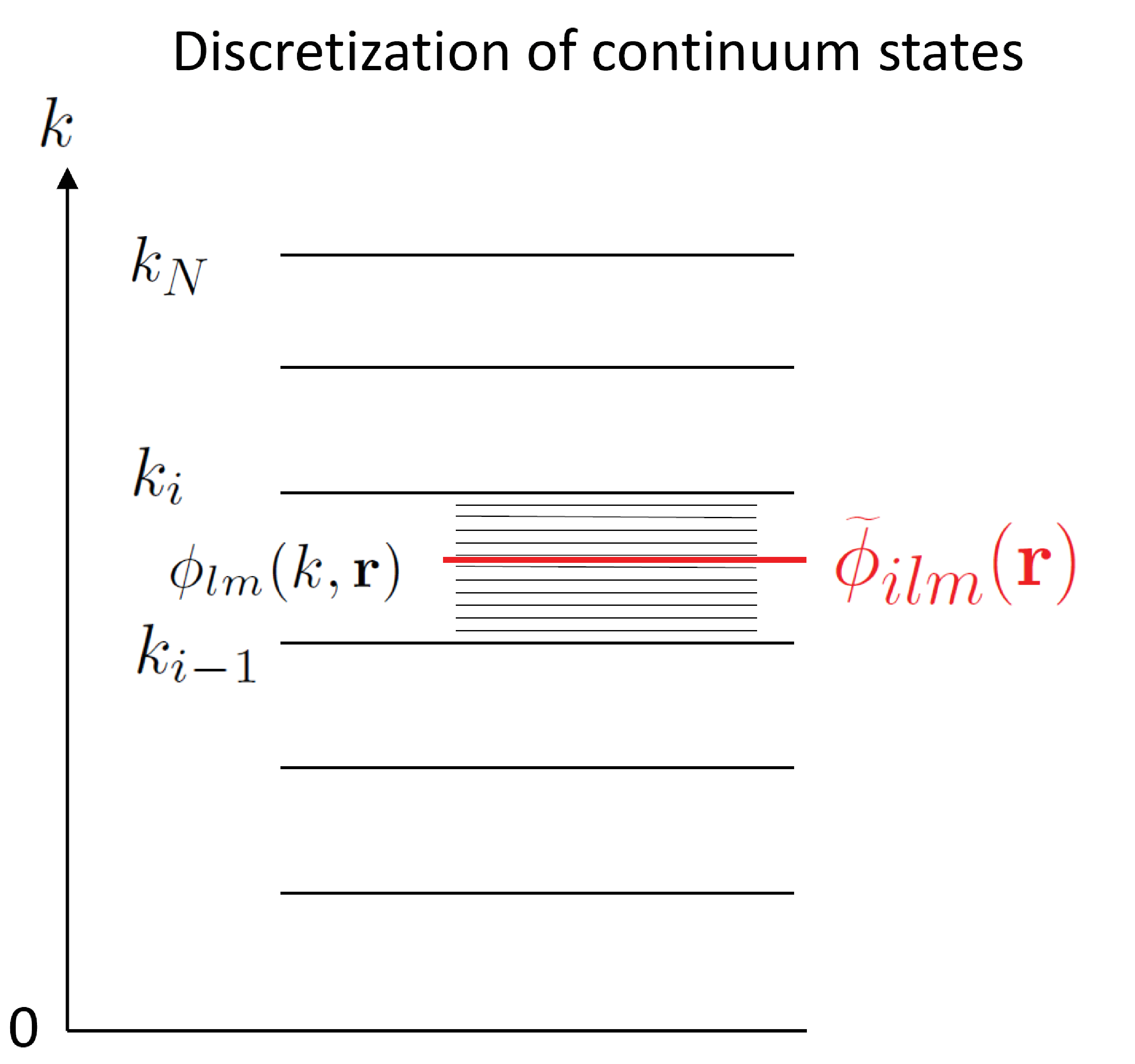,width=5.0cm,height=4.3cm}
\end{center}
\vskip -0.5cm
\caption{Schematic illustration of Eq.~(4.1) to construct the 
continuum-discretized wave function ${\widetilde \phi}_{i l m}({\bf r})$ 
by averaging the continuum wave functions $\phi_{lm}(k,{\bf r})$
in each momentum bin $\Delta k_i=k_i-k_{i-1}$. 
}
\label{fig:discretization}
\end{figure}

Because the functions $\phi_\beta(\xi_\beta)$ in the outgoing channel $\beta$
in the $T$-matrix are ortho-normalized functions of discrete states,
they discretized and ortho-normalized the $\alpha$-$\mu$ wave function 
of the $k$-continuum state, for example,  $\phi_{lm}(k, {\bf r}_5)$,  
to generate ${\widetilde \phi}_{ilm}({\bf r}_5)$ and 
energy ${\widetilde \varepsilon}_{i}$.
This discretization was performed by employing the 
procedure (Fig.~\ref{fig:discretization}):
\begin{eqnarray}
&& \!\!\!\!\!\!\! {\widetilde \phi}_{i l m}({\bf r}_5)
 =\frac{1}{\sqrt{\mathit{\Delta}k_i}}
\int_{k_{i-1}}^{k_{i}} \!\! \phi_{lm}(k,{\bf r}_5)\, dk, \;  \; \;\;
  i=1 - N, \\
\label{eq:bin}
&& \!\!\!\!\!\!\! {\widetilde \varepsilon}_{i}
=\frac{\hbar^2}{2\mu_{r_5}} {\widetilde k}_i^{\,2},  \qquad
    {\widetilde k_i}^{\,2}=\Big(\frac{k_i+k_{i-1}}{2}\Big)^2 
+ \frac{\mathit{\Delta}k_i^2}{12}. \; 
\end{eqnarray}
This has often been  used
in the continuum-discretized couple-channels (CDCC) method
for studying the projectile-breakup reactions
(for example, see review papers~\cite{Kamimura86,Austern,Yahiro12}).
On the other hand, $\phi_{nlm}({\bf r}_5)$ stands for
the bound states $(\alpha \mu)_{nlm}$.
Then, it can be said that the Ref.~\cite{Kamimura:2021msf}
investigated the reaction
\begin{eqnarray}
\label{eq:mucf-sticking1}
(dt\mu)_{J=v=0}  &\to& (\alpha \mu)_{il} + n  + 17.6 \,\mbox{MeV}  \\
\label{eq:mucf-sticking2}
      &\searrow&     (\alpha \mu)_{nl} + n  + 17.6 \,\mbox{MeV},
\end{eqnarray}
considering a precise discretization of the $\alpha$-$\mu$ continuum.

%

In our model, we replace $\Psi_{\frac{3}{2} M}^{(+)}(E)$ with 
$\Phi^{(J=v=0)}_{\frac{3}{2} M}(dt\mu)$ 
in the \mbox{$T$-matrix} elements (5.2)--(5.3)
in Ref.~[5] ---- This is a key point of the proposed model.
$\Phi^{(J=v=0)}_{\frac{3}{2} M}(dt\mu)$  is the ground-state wave function of the
$(dt\mu)$ molecule with the total angular momentum spin 3/2 and was obtained in 
Eqs.~(2.9) and (2.15) with the $d$-$t$ optical potential.  
As $\Phi^{(J=v=0)}_{\frac{3}{2} M}(dt\mu)$ 
does not explicitly include the  $\alpha n \mu$ amplitude,
the third lines of Eq.~(5.2)--(5.3) in Ref.~[5] were excluded.  
$\mathring{\Psi}_{\frac{3}{2} M}^{({\rm C})}(dt\mu) 
+ \Psi_{\frac{3}{2} M}^{({\rm N})}(dt\mu)$ that is the dominant component
of $\Psi_{\frac{3}{2} M}^{(+)}(E)$ is replaced
with $\Phi^{(J=v=0)}_{\frac{3}{2} M}(dt\mu)$.
Consequently, we obtain our $T$ matrix elements with the outgoing wave functions
in channel $c=5$:
\begin{eqnarray}
\label{eq:T5-conti}
\!{\widetilde T}^{(5)}_{il, m m_s}\!\!  = \! \langle \,
   e^{ i {\widetilde {\bf K}}_i \cdot {\bf R}_5 } \,{\widetilde \phi}_{ilm}({\bf r}_5)
       \chi_{\frac{1}{2} m_s}\!(n)\,  
    |\,  V^{({\rm T})}_{\alpha n, dt} \,| \,
 \Phi^{(J=v=0)}_{\frac{3}{2} M}(dt\mu) \,\rangle  \quad
\end{eqnarray}
for the transition to discretized states $(\alpha \mu)_{ilm}$, and
\begin{eqnarray}
\label{eq:T5-bound}
\!T^{(5)}_{nl, m m_s}\!\!  = \! \langle \,
   e^{ i {\bf K}_n \cdot {\bf R}_5 } \,\phi_{nlm}({\bf r}_5)
       \chi_{\frac{1}{2} m_s}\!(n)\,  
    |\,  V^{({\rm T})}_{\alpha n, dt} \,| \,
 \Phi^{(J=v=0)}_{\frac{3}{2} M}(dt\mu) \,\rangle  \quad
\end{eqnarray}
for the transition to  bound states $(\alpha \mu)_{nlm}$.
Here, we take ${\bf R}_4={\bf R}_3$ [5] (cf. Fig.~3).
The $(\alpha \mu)$-$n$ plane waves along ${\bf R}_5$ are denoted 
by $e^{ i {\widetilde {\bf K}}_i \cdot {\bf R}_5 }$ and 
$e^{ i {\bf K}_n \cdot {\bf R}_5 }$ in which the momentum ${\widetilde K}_i$ 
and $K_n$ are derived from energy conservation as follows
(cf. Fig.~2 and Sec.~V of Ref.~\cite{Kamimura:2021msf}):
\begin{eqnarray}
&&\!\!\!\!\!\!\!
{\widetilde E}_{i}  +
{\widetilde \varepsilon}_{i} = E_{00} + Q, \;\;\; \quad  
{\widetilde E}_{i}=\frac{\hbar^2}{2\mu_{R_5}}{\widetilde K}_i^2, \\
&&\!\!\!\!\!\!\!
E_{n} + {\varepsilon}_{n}= E_{00} + Q , 
\;\;\; \quad E_{n}=\frac{\hbar^2}{2\mu_{R_5}} K_n^2 \, , 
\label{eq:E-conservation-2} 
\label{eq:average-energy}
\end{eqnarray}
where ${\widetilde \varepsilon}_{i}$ ($\varepsilon_{n}$)  
is the energy of ${\widetilde \phi}_{ilm}({\bf r}_5)$ ($\phi_{nlm}({\bf r}_5)$)
and $E_{00} \simeq$  \mbox{3.030 keV} which is negligible 
in the present scattering problem compared with $Q=17.6$ MeV.

For the above discretization of the $\alpha$-$\mu$ continuum states
$\phi_{lm}(k,{\bf r}_5)$ , we considered $N=200$ for $l=0$ to 25,
and maximum momentum $\hbar k_N=10 {\rm MeV}/c \,
({\widetilde \varepsilon}_{N}=487$ keV) with the constant $\Delta k_i$,
which is the same as those used in Ref.~\cite{Kamimura:2021msf}.

Reaction rates of the reactions (\ref{eq:mucf-sticking1}) 
and (\ref{eq:mucf-sticking2}) are expressed as follows:
\begin{eqnarray}
  {r}_{nl}&=& {v}_{nl}
      \left( \frac{\mu_{R_5}}{2 \pi \hbar^2} \right)^2
  \times \!
   \sum _{m,m_s}
   \int \big| {T}^{(5)}_{nl,m m_s}\big|^2 \,
     {\rm d}{\widehat {\bf K}_n},   \\
{\widetilde r}_{il}&=& {v}_{il}
      \left( \frac{\mu_{R_5}}{2 \pi \hbar^2} \right)^2
  \times \!
     \sum _{m,m_s}
   \int \big|  {\widetilde T}^{(5)}_{il,m m_s}
  \big|^2 \,
     {\rm d}{\widehat {\widetilde {\bf K}}_i},
\end{eqnarray}
which is derived respectively by approximating Eqs.(5.6) and (5.7) in 
Ref.~\cite{Kamimura:2021msf} according to our proposed model.
Here, $v_{il} =\hbar {\widetilde K}_i/\mu_{R_5}$ is the velocity of the
\mbox{$(\alpha \mu)_{il}$-$n$} relative motion associated with ${\bf R}_5$,
and similarly for ${v}_{nl}=\hbar K_n/\mu_{R_5}$.
The sum of the quantum number $n$ for $r_{nl}$ and $i$ 
for ${\widetilde r}_{il}$ yields the total reaction rate 
$r^{\rm bound}_l$ for the bound states and $r^{\rm cont}_l$ for the continuum states:
\begin{eqnarray}
    r_l^{\rm bound}= \sum_n r_{nl}, \qquad  r_l^{\rm cont}= \sum_{i=1}^{N} 
{\widetilde r}_{il}.
\label{eq:rate-bound}
\end{eqnarray}

\begin{figure}[t]
\setlength{\abovecaptionskip}{0.cm}
\setlength{\belowcaptionskip}{-0.cm}
\centering 
\includegraphics[width=0.46\textwidth]{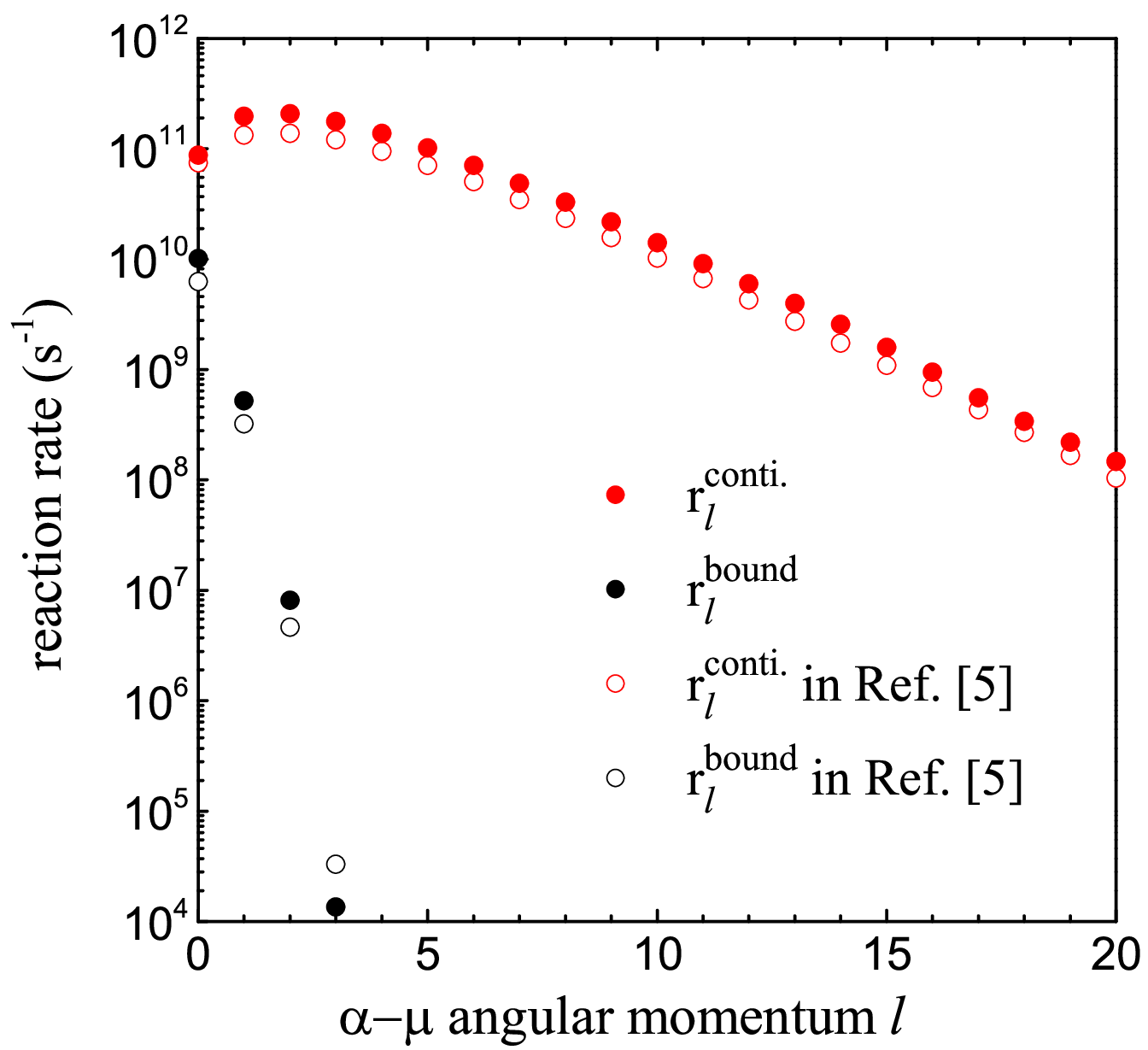}
\caption{Calculated reaction rate $r^{\rm bound}_l$ (black closed circle) 
for the bound state and $r^{\rm cont}_l$ (red closed circle) for the continuum states 
with respect to the angular momentum $l$. The potential Set B1 is used.  
The black and red  open circles denote the $r^{\rm bound}_l$ and 
the $r^{\rm cont}_l$ from Ref.~\cite{Kamimura:2021msf}, respectively.}
\label{fig:rl}
\end{figure}

In Fig.~\ref{fig:rl}, we show the calculated $r^{\rm bound}_l$ and $r^{\rm cont}_l$ 
with $l$ up to 20 under the potential Set B1 in Table~\ref{tab:vdtan}.
For comparison, we presented the $r^{\rm bound}_l$ and $r^{\rm cont}_l$ 
obtained in Ref.~\cite{Kamimura:2021msf}.
Both results have similar tendency with respect to $l$.

%

\begin{table}[t!]
\caption{Fusion rates $\lambda_{\rm f}^{(5)}$  calculated on the channel $c=5$
with Eq.~(\ref{eq:lambda-sum}). $\omega_S^0$ is 
the $\alpha$-$\mu$ sticking probability
calculated with Eq.~(\ref{eq:initial-stick}). 
See Table~\ref{tab:vdtan} for the potential sets. 
Another type of the fusion rate $\lambda_{\rm f}^{(4)}$ 
is discussed in Sec.~V on the channel $c=4$.
$\lambda_{\rm f}^{(5)}=(1.15 \pm 0.05)\times10^{12}{\rm s}^{-1}$ and
$\lambda_{\rm f}^{(4)}=(1.15 \pm 0.04)\times10^{12}{\rm s}^{-1}$, 
on average.
}
\label{tab:lambda-f45}
  \begin{tabular}{cccc}
\noalign{\vskip 0.1 true cm}
    \hline
    \hline
\noalign{\vskip 0.1 true cm}
potential set   &
$ \;\;\lambda^{(5)} \;\;$  & $\;\omega_S^0 \;$ & $\;\; \quad \quad \lambda^{(4)} \;\;$\\
\noalign{\vskip 0.1 true cm}
  & $\quad (10^{12}\,{\rm s}^{-1})\quad$ & 
$ \quad (\%)\quad $  & $\quad \quad (10^{12}\,{\rm s}^{-1})\quad$ \\
\noalign{\vskip 0.1 true cm}
    \hline
\noalign{\vskip 0.1 true cm}
   Set A1     & 1.12  & 0.942 & \quad 1.11     \\
\noalign{\vskip 0.05 true cm}
   Set A2     & 1.15  & 0.943 &\quad 1.15     \\
\noalign{\vskip 0.05 true cm}
   Set A3      & 1.11  & 0.942 &\quad 1.11    \\
\noalign{\vskip 0.05 true cm}
   Set A4     & 1.13  & 0.944  &\quad 1.12    \\
\noalign{\vskip 0.05 true cm}
\hline
\noalign{\vskip 0.05 true cm}
   Set B1    & 1.14  & 0.942   &\quad 1.13   \\
\noalign{\vskip 0.05 true cm}
   Set B2    & 1.14  & 0.940  &\quad 1.12     \\
\noalign{\vskip 0.05 true cm}
   Set B3     & 1.13  & 0.940 &\quad 1.12     \\
\noalign{\vskip 0.05 true cm}
   Set B4    & 1.14  & 0.939   &\quad 1.15    \\
\noalign{\vskip 0.05 true cm}
\hline
\noalign{\vskip 0.05 true cm}
   Set C1     & 1.14  & 0.928  &\quad 1.14   \\
\noalign{\vskip 0.05 true cm}
   Set C2     & 1.18  & 0.931  &\quad 1.19    \\
\noalign{\vskip 0.05 true cm}
   Set C3     & 1.18  & 0.945  &\quad 1.18    \\
\noalign{\vskip 0.05 true cm}
   Set C4     & 1.20  & 0.936  &\quad 1.19    \\
\noalign{\vskip 0.05 true cm}
\hline
\noalign{\vskip 0.05 true cm}
   Set D1     & 1.13  & 0.943 &\quad  1.13     \\
\noalign{\vskip 0.05 true cm}
   Set D2      & 1.13  & 0.941 &\quad  1.14    \\
\noalign{\vskip 0.05 true cm}
   Set D3    & 1.15  & 0.937 &\quad  1.13      \\
\noalign{\vskip 0.05 true cm}
   Set D4      & 1.17  & 0.930 &\quad  1.18    \\
\noalign{\vskip 0.05 true cm}
\hline
\noalign{\vskip 0.05 true cm}
   Set E1     & 1.16  &\ 0.935  &\quad  1.15    \\
\noalign{\vskip 0.05 true cm}
   Set E2    & 1.15  & 0.933 &\quad  1.14      \\
\noalign{\vskip 0.05 true cm}
   Set E3     & 1.16  & 0.938  &\quad  1.15    \\
\noalign{\vskip 0.05 true cm}
   Set E4     & 1.17  & 0.937  &\quad  1.17    \\
\noalign{\vskip 0.05 true cm}
\hline
\hline
\noalign{\vskip 0.1 true cm}
\end{tabular}
\end{table}

The total reaction rates to the $\alpha$-$\mu$ bound states and  
continuum states are defined as follows:
\begin{eqnarray}
 \lambda_{\rm f}^{\rm bound}=\sum_{l=0}^{5}\: r_l^{\rm bound}, \qquad
 \lambda_{\rm f}^{\rm cont.}=\sum_{l=0}^{20}\: r_l^{\rm cont.}.
\label{eq:lambda-cont}
\end{eqnarray}
Consequently, their sum 
\begin{eqnarray}
 \lambda^{(5)}_{\rm f}= \lambda_{\rm f}^{\rm bound}
                 +\lambda_{\rm f}^{\rm cont.}.
\label{eq:lambda-sum}
\end{eqnarray}
is the fusion rate of the $(dt\mu)_{J=v=0}$ molecule 
calculated on channel $c=5$. 

Table \ref{tab:lambda-f45} lists the fusion rates $\lambda^{(5)}_{\rm f}$ 
for all 20 different potential Sets in Table \ref{tab:vdtan}. 
All the fusion rates $\lambda^{(5)}_{\rm f}$ are within a small range of
$1.1\!-\!1.2\times10^{12}{\rm s}^{-1}$.
This proves that the choice of the potentials does not influence the fusion 
rate significantly. 
This fusion rate is consistent with $\lambda_{\rm f}^{(0)}$,
which was derived in \mbox{Sec.~II B} from the imaginary part of the complex eigenenergy
of the $(dt\mu)_{J=v=0}$ molecule.
Notably, in the proposed model, 
the fusion rate $\lambda^{(5)}_{\rm f}$ have been calculated
based on the amplitude of the outgoing $\alpha n \mu$ wave.

The initial $\alpha$-$\mu$ sticking probability $\omega_S^0$ 
is probability of the muon being captured by an $\alpha$ particle after 
the fusion happens. This is expressed by the following formula in the
same manner as in Eq.~(5.3) in Ref.~[5]:
\vspace{-0.04cm}
\begin{eqnarray}
\omega_S^0=\frac{\lambda_{\rm f}^{\rm bound}}  
            {\lambda_{\rm f}^{\rm bound} + \lambda_{\rm f}^{\rm cont.}}.
\label{eq:initial-stick}
\end{eqnarray}

The second column of Table \ref{tab:lambda-f45} presents the initial sticking 
probabilities $\omega^0_s$ for different potential sets.
The $\omega^0_s (=0.938 \pm 0.07 \%)$ are consistent 
and are close to the results ($0.91\% \sim 0.93\%$)
obtained by using  optical-potential  models
~\cite{KAMIMURA1989-BENCHMARK,Bogdanova:1981mj} and by the R-matrix 
methods~\cite{Struensee88a,
Szalewicz90,Hale93,Hu1994,Cohen1996,Jeziorski91}
considering  the $d$-$t$ nuclear interaction; 
note that our calculation is based on the 
{\it absolute} values of $\lambda_{\rm f}^{\rm bound}$ and 
$\lambda_{\rm f}^{\rm cont.}$.

However, the value of $\omega^0_s$ in Table V are $\sim \!9\%$ larger than 
$\omega^0_s\, (=\!0.857\%)$ given by Ref.~\cite{Kamimura:2021msf}
in which the coupling to the $\alpha n \mu$ channel is 
included explicitly.
This may be attributed to the following reason.
In the CC work~[5],
strong coupling between the $\alpha n \mu$ outgoing amplitude
$\Psi_{\frac{3}{2} M}^{(+)}(\alpha n \mu)$
and the nuclear $dt\mu$ amplitude 
$\Psi_{\frac{3}{2} M}^{({\rm N})}(dt\mu)$ 
is expected to enhance the contribution of the transition to
the $\alpha$-$\mu$ continuum states in Eqs.~(5.2)-(5.3) of Ref.~[5]
more than that to the $\alpha$-$\mu$ bound states, which
enhances the $\lambda_{\rm f}^{\rm cont.}$, and 
then reduces $\omega_S^0$.  However, the present model does not exhibit 
such a CC effect.

\section{Momentum and energy spectra of muon emitted by {\boldmath $\mu$}CF}

In this section, according to Step v), we calculate the 
momentum and energy spectra of the muons emitted by reaction (1.2)
and derive another type of  fusion rate of the $(dt\mu)_{J=v=0}$ molecule.
We perform  a $T$-matrix calculation by referring to 
Sec.~VI of Ref.~\cite{Kamimura:2021msf} with $(\alpha n)$-$\mu$  outgoing waves 
in channel $c=4$ (Fig.~3). 
We  discretize and ortho-normalize the wave functions of the 
$\alpha$-$n$ continuum states; for example, $\phi_{lm}(k, {\bf r}_4)$ with $l=2$,  
generating ${\widetilde \phi}_{ilm}({\bf r}_4)$
in the same manner as described in Sec.~IV. 

We begin with the $T$-matrix
expressed as Eq.~(6.2) in Ref.~\cite{Kamimura:2021msf} 
where $\Psi_{\frac{3}{2} M}^{(+)}(E)$
is the total wave function obtained using the CC Sch{\" o}dinger equation
(3.7) in Ref.~\cite{Kamimura:2021msf}. 
In the proposed  model, we replace   $\Psi_{\frac{3}{2} M}^{(+)}(E)$
with $\Phi^{(J=v=0)}_{\frac{3}{2} M}(dt\mu)$ which
is the ground-state wave function of the
$(dt\mu)$ molecule obtained from 
Eqs.~(2.9) and (2.15)
including the \mbox{$d$-$t$ optical} \mbox{potential ---- This} is the same  
key point of our model as mentioned in Sec.~IV.
Thus, in the three $T$-matrix elements in Eq.~(6.2) 
of Ref.~\cite{Kamimura:2021msf}, 
we exclude the third line because the wave function 
$\Phi^{(J=v=0)}_{\frac{3}{2} M}(dt\mu)$ 
does not have an $\alpha n \mu$ scattering amplitude, and  
replace $\mathring{\Psi}_{\frac{3}{2} M}^{({\rm C})}(dt\mu) 
+ \Psi_{\frac{3}{2} M}^{({\rm N})}(dt\mu)$
with $\Phi^{(J=v=0)}_{\frac{3}{2} M}(dt\mu)$.

We obtain   $T$-matrix elements as follows (cf. Eq.~(4.5)):
\begin{eqnarray}
\label{eq:T4-muon}
&& \!\!\!\!\!\!\!\!\!\!\!\! T^{(4)}_{il, m m_s}\!\!  = \! \langle \,
   e^{ i {\widetilde {\bf K}}_i \cdot {\bf R}_4 } \,{\widetilde \phi}_{ilm}({\bf r}_4)\,
       \chi_{\frac{1}{2} m_s}\!(n)\,  
    |\,  V^{({\rm T})}_{\alpha n, dt} \,| \,
 \Phi^{(J=v=0)}_{\frac{3}{2} M}(dt\mu) \,\rangle . \qquad \;
\end{eqnarray}
The outgoing wave is located in channel $c=4$
and  composed of the plane wave 
$e^{ i {\widetilde {\bf K}}_i \cdot {\bf R}_4 }$ and  discretized 
ortho-normalized  \mbox{$\alpha$-$n$} continuum states 
${\widetilde \phi}_{ilm}({\bf r}_4)$, 
which is constructed in the same manner as that in Eq.~(4.1) in Sec.~IV.
\begin{eqnarray}
{\widetilde \phi}_{i l m}({\bf r}_4)
 =\frac{1}{\sqrt{\mathit{\Delta}k_i}}
\int_{k_{i-1}}^{k_{i}} \!\! \phi_{lm}(k,{\bf r}_4)\, dk \, , \;
  (i=1 - N).
\label{eq:bin}
\end{eqnarray}
The average energy $\widetilde{\varepsilon}_{i}$ and momentum 
${\widetilde k}_i$ of ${\widetilde \phi}_{i l m}({\bf r_4})$ are given, 
similarly to Eq.~(4.2).
Therefore, the momentum ${\widetilde K}_i$ of the plane wave 
$e^{ i {\widetilde {\bf K}}_i \cdot {\bf R}_4 }$ is derived from the 
following energy conservation ($Q=17.6$ MeV):
\begin{eqnarray}
 {\widetilde E}_{i} 
+ {\widetilde \varepsilon}_{i}=E_{00}+Q ,  \qquad
 {\widetilde E}_{i}
=\frac{\hbar^2}{2\mu_{R_4}}{\widetilde K}_i^2.
\label{eq:conserv-muonsp}
\end{eqnarray}
%
\begin{figure}[t!]
\begin{center}
\epsfig{file=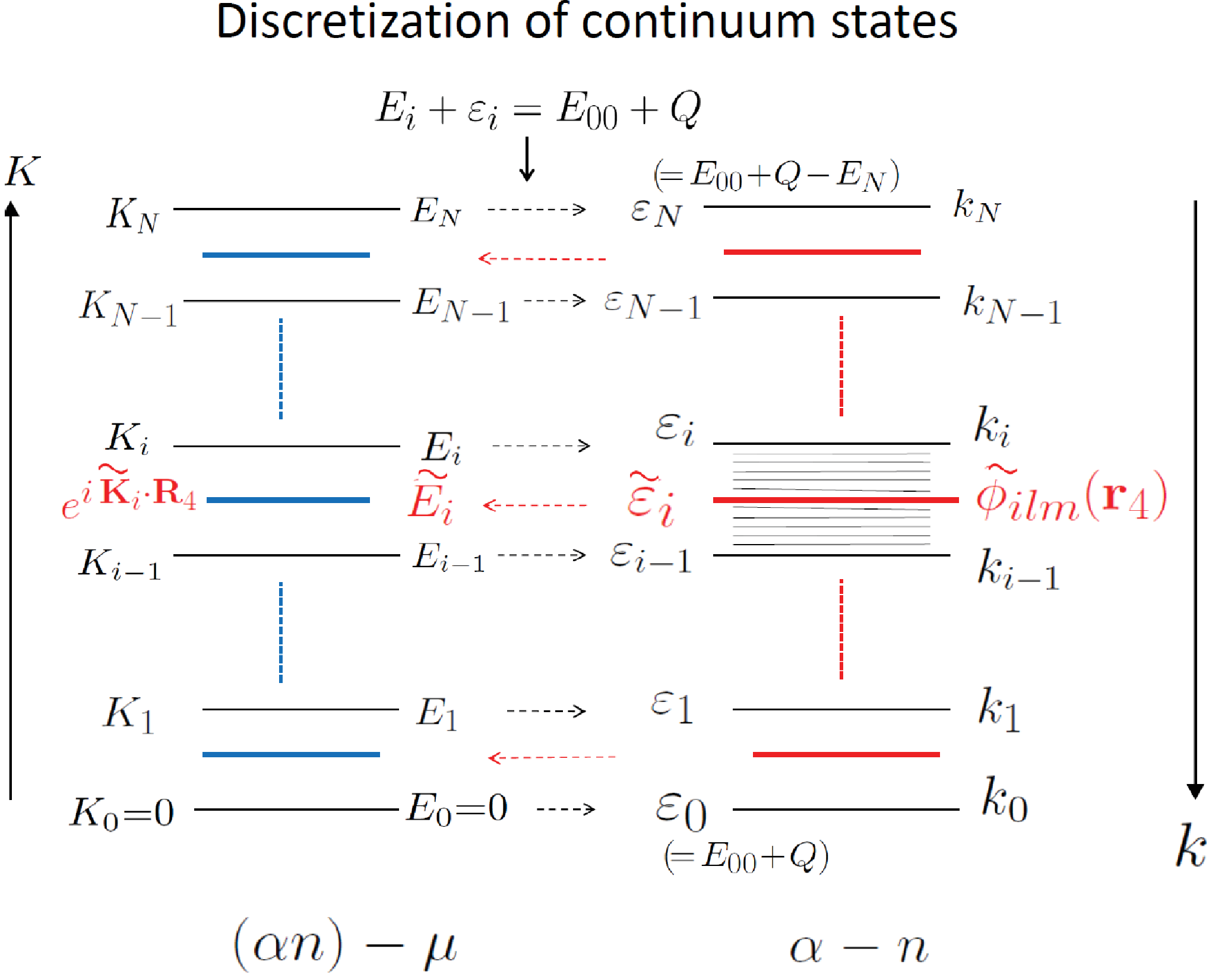,width=8.2cm,height=6.5cm}
\end{center}
\vskip -0.5cm
\caption{Schematic illustration for discretization of the 
momentum space $[K_0, K_N]$
of the ($\alpha n$)-$\mu$ relative motion 
along ${\bf R}_4$ (left half) and  that of 
the $\alpha$-$n$  relative motion 
$[k_N, k_0]$ 
along ${\bf r}_4$ (right half)
while maintaining 
$E_i + \varepsilon_i = E_{00} + Q$.
The resulting discretized \mbox{$\alpha$-$n$} continuum states 
${\widetilde \phi}_{ilm}({\bf r}_4) (i=1 - N)$ are indicated by the red lines.
The associated muon plane waves 
$e^{ i {\widetilde {\bf K}}_i \cdot {\bf R}_4 }$ are
indicated by the 
blue lines in the left half. 
This illustration is taken from Ref.~\cite{Kamimura:2021msf}.
}
\label{fig:risanka-muon-new-2}
\end{figure}

A new problem in Sec.~V is that we aim  to generate
the set $\{ {\widetilde  K}_i ; i=1-N \}$ with equal intervals
and obtain the momentum spectrum
as a smooth function of $K$ 
to determine its peak easily.
Figure~\ref{fig:risanka-muon-new-2} illustrates the manner in which
$k$-space is discretized.
Following Ref.~[5], we first assume the maximum value 
$K_N$ of  \mbox{$K$-space} $[K_0, K_N]$
with \mbox{$K_0=0$} at the left side of the figure.
We then divide the $K$-space 
\mbox{into $N$ bins ($K_i, i=0 - N$)}
with equal intervals $\mathit{\Delta}K$.
Correspondingly, we divide  the \mbox{$k$-space} $[k_N, k_0]$
on the right side of the figure to 
\mbox{$N$ bins ($k_i, i=0 - N$)}
with the energy \mbox{conservation kept as}
\begin{eqnarray}
\!\!\!\!\! E_i + \varepsilon_{i}=E_{00}+Q,  \quad 
 E_i=\frac{\hbar^2}{2\mu_{R_4}}  K_i^2,    \quad 
 \varepsilon_i=\frac{\hbar^2}{2\mu_{r_4}}  k_i^2 ,
\label{eq:conserv-muonsp-i}
\end{eqnarray}
where $K_i$ increases ($k_i$ decreases) with increase in $i$.
The bin width  \mbox{$\mathit{\Delta}k_i=|k_i-k_{i-1}|$} depends on $i$.
Subsequently, ${\widetilde \phi}_{ilm}({\bf r}_4)$ is generated 
by using Eq.~(5.2) with  energy ${\widetilde \varepsilon}_i$.
Finally, ${\widetilde E}_{i}$ is expressed
as Eq.~(5.3) as shown in Fig.~\ref{fig:risanka-muon-new-2}.

Similar to Ref.~~\cite{Kamimura:2021msf},
we consider $N=200$ setting $\hbar K_N=6.0 $\, MeV/$c \,(E_N=175$ keV)
in Fig.~\ref{fig:risanka-muon-new-2}.
This is sufficient for deriving the muon momentum spectrum
with a smooth function.

\subsection{Fusion rate of the \boldmath{$dt\mu$} molecule}

In our model, instead of Eq.~(6.4) in Ref.~[5], we can write  
the reaction rate to a continuum  discretized state $(\alpha n)_{il}$, 
\begin{equation}
  (dt\mu)_{J=v=0}  \to (\alpha n)_{il}  + \mu  + 17.6 \,\mbox{MeV}, 
\end{equation}
as follows:
\begin{eqnarray}
\label{eq:reaction-rate-Sec6}
{r}_{il}&=& {v}_{il}
      \left( \frac{\mu_{R_4}}{2 \pi \hbar^2} \right)^2
       \sum _{m,m_s}
   \int \big|  {T}^{(4)}_{il,m m_s} \big|^2 \,
     {\rm d}{\widehat {\widetilde {\bf K}}}_i \,,
\end{eqnarray}
where ${v}_{il}=\hbar {\widetilde K}_i/\mu_{R_4}$ is the 
$(\alpha n)_{il}$-$\mu$ relative velocity.

The sum of the transition rates 
\begin{eqnarray}
\lambda_{\rm f}^{(4)}=\sum_{il}\; {r}_{il}
\label{eq:fusion-rate-muon}
\end{eqnarray}
is the fusion rate of the $(dt\mu)_{J=v=0}$ molecule
using the $T$-matrix based on channel $c=4$ 
(cf. Eq.~(6.5) in Ref.~\cite{Kamimura:2021msf}).
The Contribution to $\lambda_{\rm f}^{(4)}$
from the final states ${\widetilde \phi}_{ilm}({\bf r}_4)$
with $l \neq 2$ is negligible under the present $dt$-$\alpha n$ tensor coupling
interaction.
 
The calculated fusion rates $\lambda_{\rm f}^{(4)}$
are presented in the final column of Table \ref{tab:lambda-f45}
for 20 sets of potentials listed in Table \ref{tab:vdtan}.
$\lambda_{\rm f}^{(4)}=1.1\!-\!1.2\times10^{12}{\rm s}^{-1}$ 
exhibits  minimal dependence on the potential sets and agrees 
with the result ($1.15\times10^{12}{\rm s}^{-1}$)
presented in Ref.~\cite{Kamimura:2021msf}. 
Moreover, the fusion rates $\lambda_{\rm f}^{(4)}$ 
and $\lambda_{\rm f}^{(5)}$ under the same potential set yield 
almost the same values.

\subsection{Momentum and energy spectrum of the ejected muon}

This subsection presents the muon momentum and energy spectra
derived as continuous functions of $K$ and the kinetic energy
$E$, respectively, following Sec.~VI B of Ref.~\cite{Kamimura:2021msf}.
The momentum spectra,  $r(K)$,  
is obtained by smoothing ${r}_{il}$ of Eq.~(5.6) as
\begin{eqnarray}
\!\!\!\! \lambda_{\rm f}^{(4)}= \sum_{il} 
\Big( \frac{{r}_{il}}{\mathit{\Delta}\! K} \Big)\, \mathit{\Delta}\!K
 \stackrel{\mathit{\Delta}\!K \to 0}{\longrightarrow} 
  \int_0^{K_N} \! r(K)\, {\rm d} K , 
\label{eq:muon-K-spec-1}      
\end{eqnarray}  
where the present case $\mathit{\Delta} K=0.03$ MeV is sufficiently small.
The energy distribution, ${\bar r}(E)$, is derived as follows:
\begin{eqnarray}
{\bar r}(E) \,{\rm d}E =r(K)\, {\rm d} K,\qquad
E=\hbar^2 K^2/2\mu_{{\rm R}_4}.
\label{eq:muon-E-spec-1}
\end{eqnarray}


\vskip -0.1cm  
\begin{figure}[h]
\setlength{\abovecaptionskip}{0.cm}
\setlength{\belowcaptionskip}{-0.cm}
\centering
\includegraphics[width=0.41\textwidth]{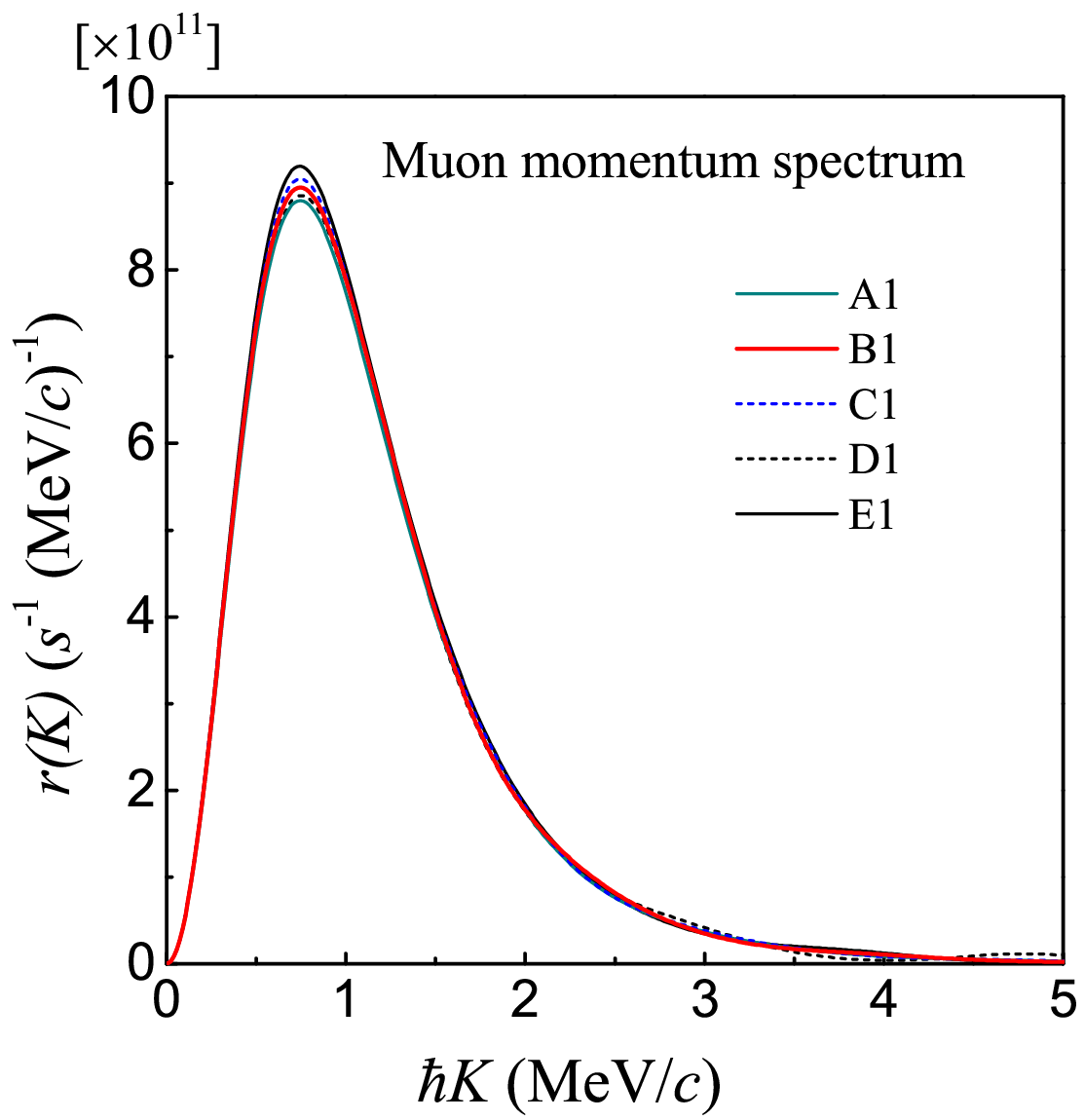}
\caption{Momentum spectrum $r(K)$ of the muon emitted by the $\mu$CF reaction (1.2),
which are calculated with Eq.~(\ref{eq:muon-K-spec-1}) using 
the potential Sets A1, B1, C1, D1, and E1 listed in Table~\ref{tab:vdtan}. 
}
\label{fig:rK}
\end{figure}
\begin{figure}[b]
\setlength{\abovecaptionskip}{0.cm}
\setlength{\belowcaptionskip}{-0.cm}
\centering
\includegraphics[width=0.41\textwidth]{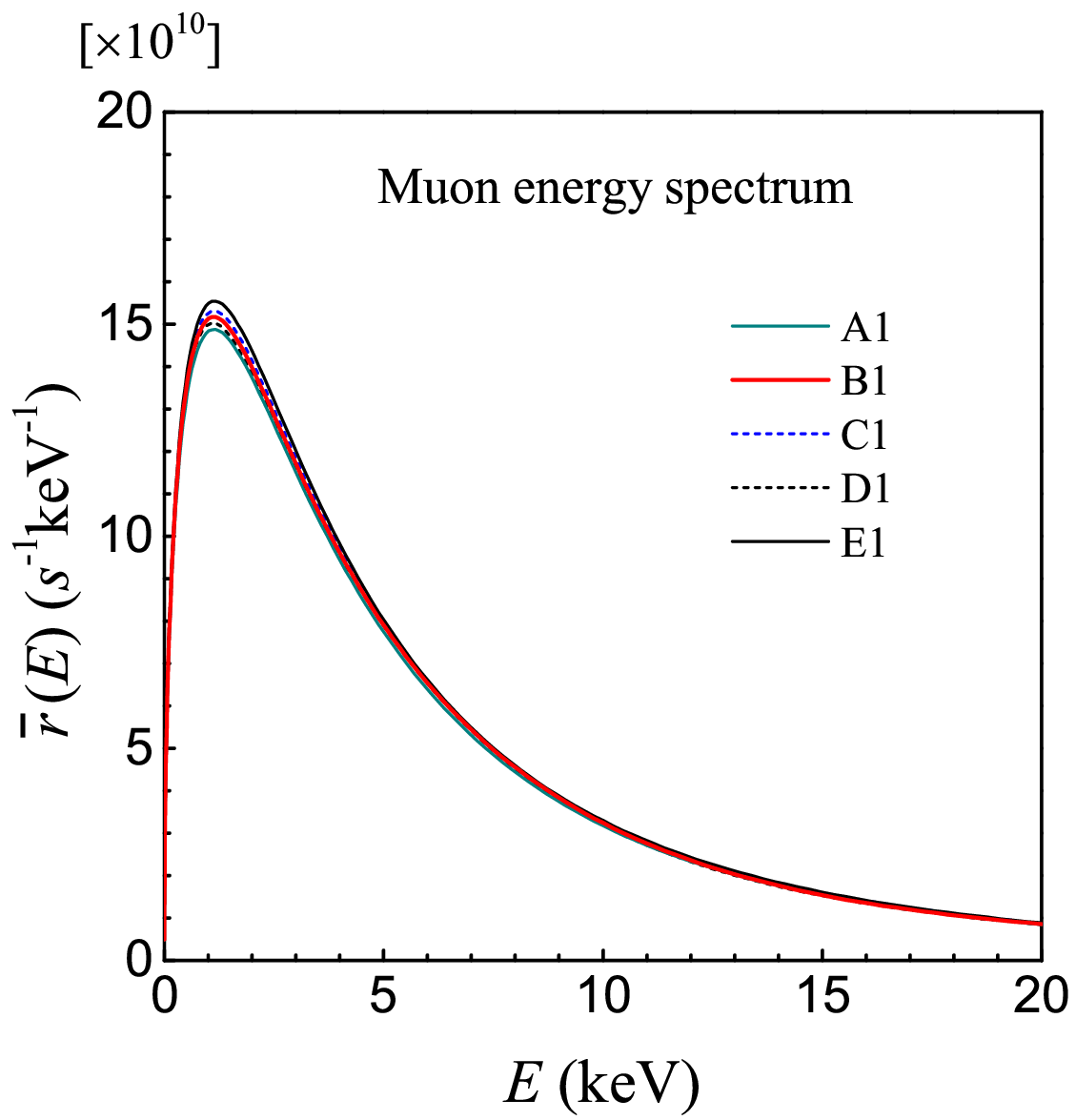}
\caption{Energy spectrum ${\bar r}(E)$ of the muon emitted by the $\mu$CF reaction
(1.2), which are calculated with Eq.~(\ref{eq:muon-E-spec-1}) using 
the potential Sets A1, B1, C1, D1, and E1 listed in Table~\ref{tab:vdtan}. 
}
\label{fig:rE}
\end{figure}
%


Figures~8 and 9 illustrate the muon momentum spectrum $r(K)$ 
and the energy spectrum ${\bar r}(E)$  
calculated using 
the potential Sets A1, B1, C1, D1, and E1 (Table~\ref{tab:vdtan}). 
The five lines  are close to each other in both the figures.
This demonstrates that our results are potentially independent of each other.
Further, we verified that the figures obtained using the potentials A3-E3
are similar to  Figs.~8 and 9. 

In Figs.~10 and 11, we compare our results for  $r(K)$ 
and ${\bar r}(E)$ (red lines using Set B1) with the results in Ref.~[5] (black 
lines) taken from Figs.~14 and 15 using Eqs.~(6.9) and (6.11), 
%
\begin{figure}[h]
\setlength{\abovecaptionskip}{0.cm}
\setlength{\belowcaptionskip}{-0.cm}
\vskip 0.2cm
\centering
\includegraphics[width=0.41\textwidth]{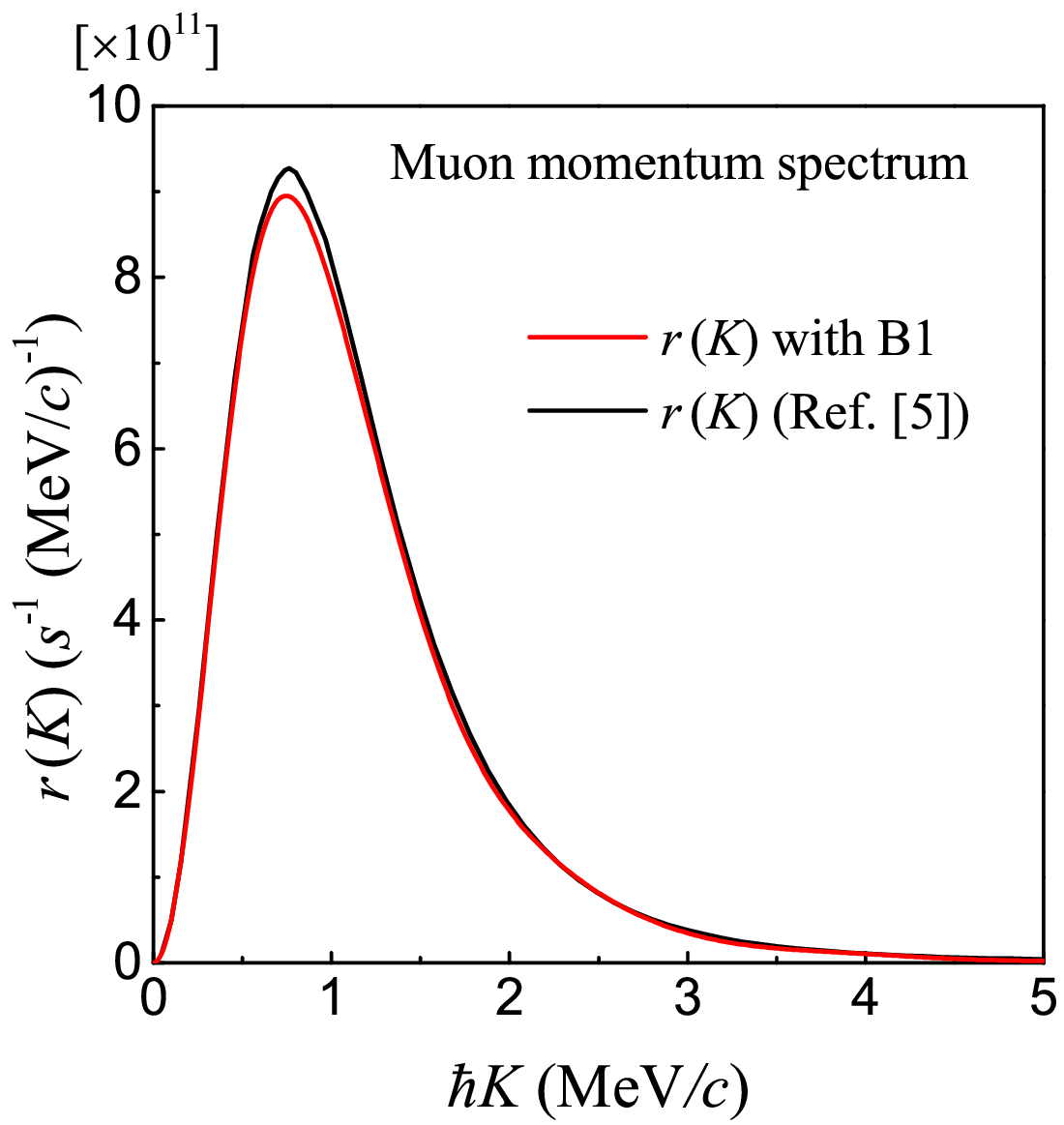}
\caption{Momentum spectrum $r(K)$ of the ejected muon 
calculated with the potential Set B1 
in Fig.~\ref{fig:rK}, which is compared with the momentum spectrum $r(K)$
presented in Fig.~14 of Ref.~\cite{Kamimura:2021msf}.
}
\label{fig:rK-2}
\end{figure}
\begin{figure}[b]
\setlength{\abovecaptionskip}{0.cm}
\setlength{\belowcaptionskip}{-0.cm}
\centering
\includegraphics[width=0.41\textwidth]{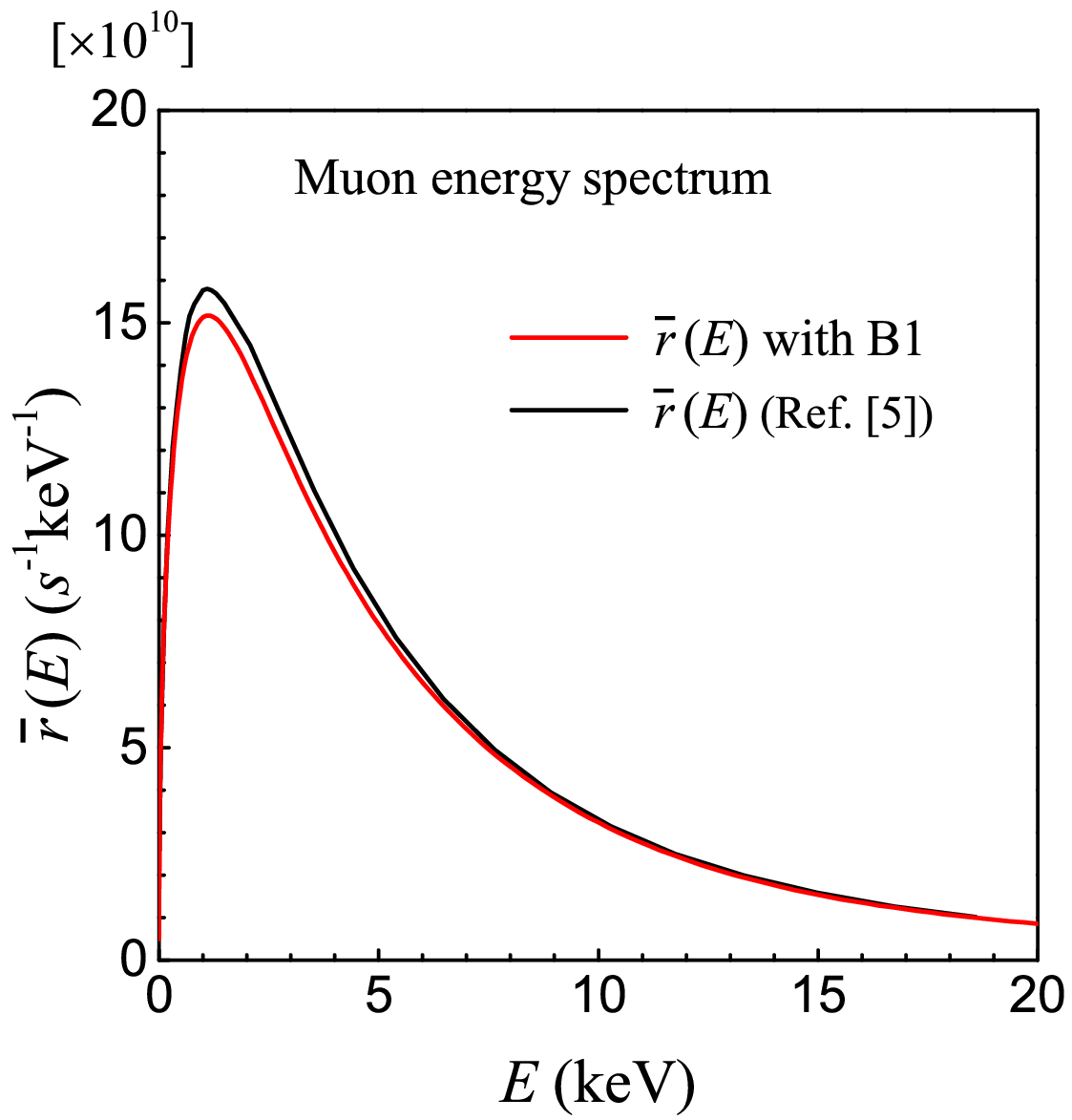}
\caption{Energy spectrum ${\bar r}(E)$ of the ejected 
muon calculated with the potential Set B1 
in Fig.~\ref{fig:rE}, which is compared with the energy spectrum ${\bar r}(E)$ 
presented in Fig.~15 of Ref.~\cite{Kamimura:2021msf}.
}
\label{fig:rE-2}
\end{figure}
%

\noindent
respectively. Our results agree well with those reported in 
Ref.~[5].The use of any other potential set also reproduces the result of Ref.~[5]
as is understood from Figs.~8 and 9.

In Table.~\ref{tab:av-f4}, the peak and average energies 
of the muon energy spectrum ${\bar r}(E)$  
and those reported in Ref.~\cite{Kamimura:2021msf} are compared.
The peak energy is \,located at\, $E \sim1.1\!$ keV  
both in our \mbox{$T$-matrix} model
and in Ref.~\cite{Kamimura:2021msf}. 
The average energy of \mbox{8.9 keV (Set B1)}, is also consistent
with the result (9.5 keV) from Ref.~\cite{Kamimura:2021msf}. 
This large average energy is caused by the long
high-energy tail of the energy spectrum as evident in Fig.~\ref{fig:rE-2}. 
Therefore, it can be concluded that 
muons with peak energy of \mbox{$\!\sim \! 1$ keV} and  
the average energy of  \mbox{$\!\sim \!10$ keV} are emitted 
by  $dt\mu$ fusion, which
is the same as that in Ref.~\cite{Kamimura:2021msf}. 
These results will be helpful to the ongoing experimental project 
for generate an ultra-slow negative muon beam 
using $\mu$CF for various applications.

\begin{table}[t] 
\caption{Property of the muon energy spectrum ${\bar r}(E)$ 
by the present result
with the use of potential Set B1 (Fig.~11). The case of ${\bar r}(E)$  
given by Ref.~\cite{Kamimura:2021msf} is also listed for comparison.
}
\begin{center}
\begin{tabular}{lcccccc}
\noalign{\vskip -0.2true cm}
\hline \hline
\noalign{\vskip 0.1 true cm}
Muon energy spectrum   & $ \;$ &  Peak  &   &
Average    &  & Peak   \\
\noalign{\vskip -0.1true cm}
  &  &   energy &   &  energy   &  &  strength  \\
\noalign{\vskip 0.01true cm}
  &  &  (keV) & $\;$  &  (keV)   &  &
    $({\rm s}\cdot{\rm keV})^{-1}$  \\
\noalign{\vskip 0.1true cm}
\hline
\noalign{\vskip 0.1true cm}
Present, ${\bar r}(E)$ with B1 &  & 1.1 &   &  8.9   &  & $1.54 \times 10^{11}$ \\
\noalign{\vskip 0.1true cm}
${\bar r}(E)$ (Ref.~\cite{Kamimura:2021msf}) 
  &  &    1.1 &   &  9.5   &  & $1.60 \times 10^{11}$ \\
\noalign{\vskip 0.1 true cm}
\hline
\hline
\noalign{\vskip -0.3 true cm}
\end{tabular}
\label{tab:av-f4}
\end{center}
\end{table}
%

\section{Conclusion}
\label{sec:conclusion}

Recently, a comprehensive study of the $\mu$CF reaction 
(1.2) was performed in Ref.~\cite{Kamimura:2021msf} 
by solving two coupled channel (CC) Schr\"odinger equations
for the reactions (1.1) and (1.2).
In the present study, we have proposed a considerably more tractable $T$-matrix
model to simulate the  CC model~\cite{Kamimura:2021msf} by considering
scenarios i) -- v):

i) We  reproduced the low-energy cross sections of  reaction (1.1)
by using an optical-potential (OP) model (Fig.~2 and Table I).
The effect of the $\alpha$-$n$ channel is considered to be \mbox{included}
using the imaginary part of the optical potential.

ii) The exact $T$-matrix 
for the reaction (1.1) was approximated by 
replacing the exact CC wave function  with the OP-model 
wave function obtained  in i).
The cross section of the reaction (1.1) was expressed by the simple 
closed form (3.10) based on our model and reproduced the observed 
cross section by  properly \mbox{selecting} the $dt$-$\alpha n$ coupling potential 
(Fig.~4 and \mbox{Table~IV)}.

iii) We calculated 
the $(dt\mu)_{J=v=0}$ molecular wave function including the OP-model
$d$-$t$ potential determined in  i).

iv) We approximated the full $T$-matrix in Ref.~[5] for the reaction (1.2)
as follows: 
We replaced the CC wave function for this reaction with the
muonic $(dt\mu)_{J=v=0}$ wave function \mbox{obtained} in iii).

v) Using the approximated $T$-matrix in iv), we calculated the reaction rates 
to the $\alpha$-$\mu$ continuum and the bound states,
fusion rates of reaction (1.2),  \mbox{$\alpha$-$\mu$} sticking probability,
and  momentum and energy spectra of  muons emitted by
reaction (1.2).  Most of the results obtained in Ref.~[5] were well reproduced.

All the calculated results were insensitive to  20 sets of 
$d$-$t$ potential and  $dt$-$\alpha n$ coupling potential. 
In practical calculations,
the use of only a few sets is sufficient.

Thus, the proposed tractable $T$-matrix model was constructed, such that
it reproduced most of the results obtained in Ref.~\cite{Kamimura:2021msf}.
This model is applicable to other $\mu$CF systems such as $(dd\mu)$,
$(tt\mu)$, $(dt\mu)^*$, and $(dd\mu)^*$.

%
\begin{acknowledgments}
The authors would like to thank Prof.~Y.~Kino and Dr.~T.~Yamashita for their 
valuable discussions in the study. This work is supported by
the Grant-in-Aid for Scientific Research on Innovative Areas,
``Toward new frontiers: Encounter and synergy of state-of-the-art
astronomical detectors and exotic quantum beams", 
JSPS KAKENHI Grant Number JP18H05461.
This work is also supported by Natural Science Foundation of 
Jiangsu Province (Grant no. BK20220122), and National Natural Science Foundation 
of China (Grant no. 12233002). The main computing was conducted 
at the Nanjing Nengmao Space Cloud Computing Technology Co., Ltd.
\end{acknowledgments}


\end{document}